\documentclass[12pt]{article}
\usepackage{amsmath,amssymb,amsthm,latexsym}
\usepackage{stmaryrd,wasysym,upgreek,mathrsfs,dsfont}
\usepackage[english]{babel}
\usepackage{graphicx,color}
\usepackage{amscd}

\usepackage[pdftex]{hyperref}

\newcommand{\be}{\begin{equation}}
\newcommand{\bee}{\begin{equation}}
\newcommand{\ee}{\end{equation}}
\newcommand{\bea}{\begin{eqnarray}}
\newcommand{\eea}{\end{eqnarray}}

\newcommand{\cG}{{\cal{G}}}

\newcommand{\cJ}{{\cal{J}}}
\begin{document}

\title{The Tensor  Track, III}


\author{Vincent Rivasseau\\
Laboratoire de Physique Th\'eorique, CNRS UMR 8627,\\
Universit\'e Paris XI,  F-91405 Orsay Cedex\\
E-mail: rivass@th.u-psud.fr}

\maketitle
\begin{abstract}

We provide an informal up-to-date review of the tensor track approach 
to quantum gravity. In a long introduction we describe in simple terms the motivations for this approach. 
Then the many recent advances are summarized, with emphasis on some points (Gromov-Hausdorff limit, Loop vertex expansion, Osterwalder-Schrader positivity...)
which, while important for the tensor track program, are not detailed in the usual quantum gravity literature. We list open questions in the conclusion and 
provide a rather extended bibliography.

\end{abstract}

\section{Introduction}

The tensor track \cite{Rivasseau:2011hm,Rivasseau:2012yp}
is an attempt to quantize gravity born at the crossroad of several closely related approaches, most notably group field theory (GFT) 
\cite{boulatov,laurentgft,Oriti:2011jm,Krajewski:2012aw,Baratin:2010wi,Baratin:2011aa}, which is the 
second-quantized, field-theoretic version \cite{Oriti:2013aqa} of loop quantum gravity \cite{Rovelli}, non commutative quantum field theory (NCQFT) 
\cite{Douglas:2001ba,Rivasseau:2007ab}, random matrix models \cite{Mehta,Di Francesco:1993nw} and (causal) dynamical triangulations \cite{ambjorn-book,scratch,Ambjorn:2013apa}. Based on random tensors and their 
recently discovered 1/N expansion \cite{Gurau:2011kk,Gur3,GurRiv,Gur4,Gurau:2011xp}\footnote{The theory
deals with {\emph{a priori unsymmetrized}} tensors; it uses a combinatorial device called color
to label their indices, and is born out of an attempt to simplify and improve group field theory,
hence its initial name of {\emph{colored}} group field theory \cite{color,Geloun:2009pe}.}, it proposes to quantize gravity as a quantum field theory (QFT) \emph{of}
space-time, not \emph{on} space-time, and suggests that our physical space-time emerges\footnote{The concept of \emph{emergent space-time}
has been attracting increasing attention in the recent years among physicists \cite{Seiberg:2006wf,Oriti:2006ar,Oriti:2007qd,Konopka:2008hp,Sindoni:2011ej}. Like quantum mechanics, 
it implies a challenging revision of our traditional representation of reality.} 
as a continuum limit of tensor-type theories through a cascade of phase transitions. Let us start 
by summarizing the motivations to use random tensors to quantize gravity.

In Feynman's powerful functional integral point of view, randomizing metric spaces
is the natural mathematical translation of quantizing gravity. 
What do we mean by a random metric space? It is not just a fractal, as ordinary fractal sets are deterministic. 
Random metric spaces should rather be non-trivial probability measures on the ``space of all metric spaces". But what is the latter? An excellent candidate
is the Gromov-Hausdorff (GH) space \cite{gromov}, which is the set of all isometry classes of compact spaces\footnote{The restriction to compact spaces is
there for mathematical convenience; we may never know if our universe is ultimately "compact" or not.}. 
Indeed equipped with the Gromov-Hausdorff distance, it is a 
metrizable complete separable space. Hence in that space convergence can be studied through sequences and
probability measures can be naturally defined with respect to the canonical Borelian $\sigma$-algebra.

The first and most basic tool to analyze Feynman's functional integral is its perturbative expansion in Feynman graphs. Graphs neatly encode
the two dual physical concepts of propagation (edges) and interaction (vertices). They are 
canonically equipped with a metric, namely the graph distance. Interesting phases 
may therefore correspond to GH limits of ensembles of Feynman graphs
when the average number of vertices becomes infinite.

A QFT with no initial reference to any underlying space-time is often called zero-dimensional. 
This terminology reflects our difficulty in getting rid of the space-time picture.
In such a QFT the field ``depends on a single space-time point". 
Erasing that dubious point, we could call such QFTs \emph{combinatorial}.

A necessary minimal ingredient to make combinatorial QFTs physically interesting is 
to consider (at least) one (complex) field\footnote{The restriction to a \emph{single complex} field is again for simplicity and not essential.} $(\phi_j, \bar\phi_j )$ with a huge number $N$ of \emph{components} $j=1, \cdots N$. Indeed the number of components of a (zero-dimensional) field should
correspond to the number of degrees of freedom of the physical system quantized. Our universe has perhaps
$10^{130}$ degrees of freedom or more\footnote{Excluding matter and restricting to pure gravitational degrees of freedom
does not reduce significantly this number.}! 

Although nature might give us some partial hints, it seems very unlikely that experiments in the next decades could truly decide
between the many competing theories of quantum gravity. Until this eventually happens, 
simplicity and mathematical consistency are rational criteria to select  and to study in depth the most promising candidates. 
In this respect a successful purely combinatorial QFT of space-time and gravity would be very appealing\footnote{
It would be indeed appealing for a fundamental theory to require little or no complicated mathematical structures at the start.
Among such theories, called "austere",  let us mention (beyond random tensors) at least dynamical 
triangulations and causal sets \cite{causalsets}.}. But several harsh conditions have to be met to qualify as successful: 
the (non-perturbative) continuum limit should be under control, universal in a certain 
sense, and should lead to a large effective space-time like our own, governed by classical general relativity\footnote{It would be nice if the formalism led also to emerging matter fields like those of the standard model, but this unification is not a priori \emph{required} for a successful quantum theory of gravity.}.
Of course we are not there yet, but we can take comfort in the fact that such a program
has essentially already been achieved for two dimensional quantum gravity. Although 
the mathematics has still to be established in greater detail, there is indeed by now some general agreement 
that purely combinatorial QFTs of the random matrix type, with their ribbon Feynman graphs dual to triangulations of surfaces,
are able to quantize two dimensional gravity and to provide a description complementary but ultimately equivalent to the 
more traditional continuum picture \cite{Di Francesco:1993nw}. 

The behavior of combinatorial QFT's with a large number $N$ of components is always studied 
through an expansion in powers of $1/N$. Until recently, physicists knew only two kinds of such $1/N$ expansions, the first one governing vector models and the second
one governing matrix models. The leading Feynman graphs for these two expansions, namely plane trees and planar graphs,
correspond to two classes of graphs which mathematicians had been able 
to count exactly, and more recently to explicitly map to simple combinations of Brownian-type processes.  Using these maps, they succeeded to define two interesting 
non-trivial probabilistic measures on the GH space, namely the continuous random tree (CRT) of Aldous \cite{aldous} (also called branched polymers in physical context)
and the Brownian (two-dimensional) sphere of Le Gall and Miermont \cite{LeGall,LeGallMiermont,Miermont}\footnote{Higher genus analogs of the Brownian sphere exist, 
but we do not discuss them here for simplicity.}.  These non-trivial random spaces
occur as GH-limits of the corresponding equidistributed Feynman graphs (after natural rescaling of the graph distance). They are quite non-trivial:
the CRT has Hausdorff dimension 2 and spectral dimension 4/3; the Brownian sphere has Hausdorff dimension 4 and expected spectral dimension 2.
They are also universal in a certain precise sense.

Once we orient ourselves towards the search of a combinatorial QFT for higher dimensional quantum gravity as well, 
the obvious step is to build on the success of vector and matrix models and to generalize them to higher rank tensor models. This idea indeed appeared shortly after
the initial success of matrix models for 2d gravity \cite{ADT1,ADT2,sasa,gross}. It was fully understood at that time that tensor models of rank three or more generate 
Feynman graphs dual to triangulations of spaces of dimension three or more. It was also understood that in any dimension $d \ge 3$
such tensor models naturally ponder these triangulations exactly 
with the discretized Einstein-Hilbert action with cosmological constant, without any other additional continuous data needed \cite{ambjorn}.
\footnote{This fact, well-known among specialists, seems perhaps not sufficiently appreciated in larger mathematics or theoretical physics circles.}.
Surprisingly perhaps, what was not identified in this early phase of tensor models was the class of invariant
interactions to consider, and the associated $1/N$ expansion. Let us pause for a while to explain this point.

A class of combinatorial QFTs is defined by the choice of an interesting propagator 
and of an interesting set of interactions between the many components of the field.
There is always an obvious choice for a \emph{canonical} propagator, namely the inverse of the quadratic form $\sum_j \bar \phi_j \phi_j$ which pairs 
components together as independent fields. Mathematically this quadratic form is a scalar product. It means that the complex field
takes values in a Hilbert space of large dimension $N$ and that the free theory is invariant under the $U(N)$
group of unitary transformations. Remark however that such a propagator is also quite trivial: being the inverse of a mass term, it misses
the essential characteristic of an ordinary QFT propagator, namely \emph{propagation}. It also lacks the associated 
physical notion of \emph{scales}. We shall return below to this point.

The next choice, less obvious but essential, is to decide which class of interactions to consider. 
It corresponds to the choice of an abstract combinatorial generalization of \emph{locality}, the characteristic property of interactions.
In ordinary QFT, interactions are products of fields \emph{at the same point}.
It is here that the combinatorial difference between vector and matrix models appears.  
Indeed, an $N_1$ by $N_2$ rectangular matrix is also a particular case of a vector, 
but in an $N= N_1 \times N_2$ dimensional Hilbert space. The canonical propagators of vectors and matrix models 
are identical. But vector model \emph{interactions} have the same full $U(N) = U(N_1 \times N_2)$, hence $(N_1^2N_2^2)$-dimensional invariance
than the free theory, whereas
matrix model interactions have only $U(N_1) \otimes U(N_2)$, hence
$(N_1^2 + N_2^2)$-dimensional invariance. This reduced invariance corresponds to independence of the choice of an orthonormal basis 
in the source and target spaces of the matrix. Hence we might consider matrix models as vector models
with a particular breaking of the vector symmetry. This breaking also means that a richer set of interactions (invariant under the restricted, broken symmetry) 
becomes available to create interacting models. Although vector models have no \emph{connected}\footnote{Here \emph{connected}
means that the graph representing the pairing of indices is connected. Polynomial invariants are linear combinations of products of connected invariants. Hence 
vector models can have interactions, but which are not \emph{connected} invariants. This is why they can be split using intermediate scalar fields, see \eqref{spliscalar}.}
invariant interactions (beyond the quadratic term), complex matrix models have exactly \emph{one} connected invariant interaction
at every (even) degree, namely $Tr (MM^\dagger)^p$.

This observation guides us to the natural class of interactions for tensor models. It simply corresponds
to a further breaking of the combinatorial symmetry. Rank $D$ tensors are vectors in a $N_1 \times \cdots \times N_D$ space, but should have only 
a tensorial $U(N_1) \otimes \cdots \otimes U(N_D)$ invariance. This invariance corresponds to independence of interactions
with respect to independent change of orthonormal basis in each of the spaces of the tensor product. In rank three or more it leads to 
a much richer set of interactions than for matrices. These interactions are in one to one correspondence with bipartite $D$-regular edge-colored graphs 
\cite{Gurau:2011tj,Bonzom:2012hw} and for $D \ge 3$ their number $Z^c_D (n)$ 
on $n$ vertices and $n$ anti-vertices increases fast with $n$ \cite{Geloun:2013kta}:
\bea
Z^c_1 (n) &=& 1, 0, 0, 0 , 0, ...  \nonumber \\  
Z^c_2 (n) &=& 1, 1,1,1,1, 1, 1, ... \nonumber \\  
Z^c_3 (n) &=& 1, 3, 7, 26, 97, 624, 4163, ... \nonumber \\  
Z^c_4 (n) &=& 1, 7, 41, 604, 13753, ... \nonumber
\eea

Tensor models  with canonical trivial propagator and a finite set of such invariant interactions 
have a perturbative expansion indexed by Feynman graphs which, under expansion of the internal structure of their bipartite $D$-regular edge-colored vertices,
become bipartite $(D+1)$-regular edge-colored graphs. They typically
admit an associated $1/N$ expansion whose leading graphs, called melons \cite{Bonzom:2011zz}, 
are in one-to one correspondence with $(D+1)$-ary trees\footnote{This statement in fact requires that the
interactions of these models are themselves melonic; see \cite{Bonzom:2012wa} for what can happen if this is not the case.}. This is essentially the breakthrough that revived the tensor program.

The tensorial $1/N$ expansion is governed by a number, 
the degree of the triangulation \cite{Gur3}, which is \emph{not} a topological invariant. This new feature is \emph{good}.
Indeed, although the problem of counting (or even exponentially bounding) and summing over all triangulations of three or four dimensional spaces within a fixed
topology (e.g. the sphere) may be hopelessly difficult, the problem of quantizing gravity in these dimensions may be \emph{easier}! Summing
triangulations dual to the Feynman graphs of tensor models pondered by their 1/N expansion factors, that is 
triangulations pondered by their natural discretized Einstein-Hilbert action, 
leads to a completely different program than summing them with uniform factors in a given topological class.  
Solving analytically for instance some finite multiple scaling limit of tensor models 
might be doable and hopefully might lead to a computable theory of quantum gravity in
three or four dimensions without necessarily solving the problem of counting or 
summing exactly all uniformly pondered triangulations of the three or four dimensional topological sphere or of other manifolds.

The first step in this direction, namely the study of single scaling in tensor models, was performed in \cite{Bonzom:2011zz}.
It was found that single scaling in tensor models, in which only melons survive, generates a "protospace" which is a CRT. It was 
also rigorously proved in the sense of the GH limit and of the Hausdorff and spectral dimensions in \cite{Gurau:2013cbh}. This
confirms the theory and numerical simulations of Euclidean dynamic triangulations \cite{ambjorn-book}. We 
do not consider this fact particularly discouraging. Since the CRT is the limit of vector models, it might be considered 
as a kind of quantum gravity in one dimension. Hence it is not so surprising to meet first this random space on the road 
towards higher dimensional quantum gravity. The main reason for reviving the search for more interesting infrared phases
is that the random tensor models provide an analytic approach with a new tool, the $1/N$
expansion, to explore beyond this branched polymer phase.

Recently a second step has been taken. The double scaling limit of tensor models has been performed 
\cite{GurauSchaeffer,Dartois:2013sra}. It is led (in dimensions smaller than 6)
by a larger family of graphs than melons. These graphs have tree like structure decorated with loops
but these loops are still too short to change the universality class and move the theory beyond the CRT. 
Again this is not too discouraging. It just indicates that emergence of a smoother continuous space-time cannot 
yet happen at double scaling, but requires more steps. As long as the leading graphs for such multiple scalings are essentially trees with short loops, they are good starting points for further scalings. This multiple scaling process should create increasing densities of loops linking the vertices of the initial CRT tree and can be hopefully continued until a non-CRT phase is found. A universe like our own, with local dimension 4 and the local degrees of freedom of general relativity could therefore emerge somewhere along this road.

At this point one could object that in matrix models the $1/N$ limit leads directly to the 2d Brownian sphere without 
the intermediate step of the CRT. But we know that the 2d Brownian sphere is best understood as a CRT 
equipped with a simple Brownian process, namely the additional labels of the Schaeffer map. 
These labels represent shortcuts on the tree and lead to a qualitatively different random space; 
but one can consider that 2d quantum gravity indeed is built on a hidden CRT.
The tensor $1/N$ expansion includes a richer structure of sub-leading terms than the vector and matrix $1/N$ expansions, and it may have
in some way lifted the degeneracy
that leads directly from matrix models to the Brownian sphere. 
Tensors in fact include matrix models in several different ways. Hence they certainly can generate 2d gravity, at least 
in a sufficiently anisotropic regime (e.g. $N_1=N_2 >>  \cdots N_D$) (see \cite{Gurau:2011sk} for an example of such a limit). 
The real challenge for tensor models is therefore to go beyond the 2d Brownian sphere rather than beyond the CRT, and the richer structure
of sub-leading terms in the tensorial $1/N$ expansion may allow for that.

This scenario in many steps for space-time emergence, starting from no space at all and creating first a branched polymer phase, then
more and more loop shortcuts leading to a smoother random space, remains of course a hypothesis to be explored in depth. Nevertheless it is suggested 
by the richer structure of the tensor $1/N$ expansion and can be considered as a new insight. This insight is fully consistent with the recent 
arguments  coming from different approaches to quantum gravity which point towards space-time appearing lower dimensional
at high ultraviolet scales, and higher dimensional in the infrared  \cite{Carlip:2009kf,Stojkovic}. This is in a way the exact opposite of the 
Kaluza-Klein (and standard superstring) paradigm that additional small compact dimensions should appear at high energy.

In short, quantum gravity, like the rest of physics, probably 
requires many intermediate descriptions between the fundamental very symmetric ultraviolet theory and 
its effective infrared products. 

Even if an \emph{ab initio} fully combinatorial approach may be a correct way to model quantum gravity 
in the extreme ultraviolet regime, the essential goal is therefore to connect it convincingly to more traditional 
effective pictures. This is a complicated task and the critical part of the tensor track program. 

Along the road from the combinatorial aspects of tensor models to more effective models lies in particular the important physical issues of
how and when to introduce a particular background topology and continuous coordinates, and the related issue of diffeomorphism invariance 
with respect to these coordinates\footnote{Diffeomorphism invariance of 
classical gravity makes desirable to quantize gravity in a \emph{background independent way}. 
This idea is probably best realized if space-time is an emergent reality and if the underlying
``pregeometric" theory does not even depend on any particular preferred topological manifold. See \cite{Baratin:2011tg} for a discussion of 
diffeomorphism invariance in the group field theory context.}. 
Whenever we try to predict the influence of quantum gravity (ie randomizing the metric) on the behavior of matter and of
classical gravity, we have to formulate such physical statements with respect to a given 
(local) \emph{fixed coordinate system}.

In 2d quantum gravity the analog problem consists in connecting the combinatorial discrete picture of triangulations
leading to the Brownian sphere in the GH limit to the continuous Liouville picture, which is a better way to understand the behavior of 
conformally invariant matter fields \cite{Polyakov1981207,Friedan:1982is}. Here the main physical
statement is that the presence of quantum gravity in two dimensions modifies the critical behavior of these fields 
through the KPZ relations \cite{KPZ:1988,David1988a,David1988b,DistlerKawai,Duplantier,Duplantier:2009np,DuplantierSheffield}. 

A recent progress along these lines \cite{davideynard} analyzes the (infinite, continuous) family of planar Delaunay triangulations
corresponding to a fixed planar combinatorial triangulation. It relates the parametrization of this family through edge-angle variables 
to the parametrization in terms of the complex coordinates of the vertices. The Jacobian of this coordinate change is identified with a discretized version of the
Faddeev-Popov determinant for the conformal gauge fixing for a 2d metric, which itself leads to the Liouville continuous action and to the computation of conformal anomalies. 
Therefore the discretized graphs equipped with continuous edge-angle variables 
sit somewhere in between the combinatorial and the continuum Liouville picture.

This prefigures what one could also hope for in higher dimensions: combinatorial triangulations 
equipped with additional continuous variables, such as dihedral angles integrated on certain 
domains to represent curvature, would help to bridge the gap between the purely combinatorial description of quantum gravity
and a continuous formalism\footnote{Remark that Delaunay graphs dual to Voronoi triangulations exist in any dimension, but they are more difficult to parametrize.}.
In three dimensions we would like such intermediate steps to bring us closer to the topological field theory formulation either as $BF$ or as Chern-Simons theory. 
In four dimensions this continuous formulation should ultimately become general relativity, which can be considered as a topological $BF$ theory 
modified by Plebanski simplicity constraints on $B$ to free the local degrees of freedom of gravitational waves
 \cite{Plebanski,Holst,Krasnov,Gielen:2010cu}.

Group field theory fits this description for such an intermediate picture.  It uses discretized triangulations, but equipped with additional Lie-group data 
which in two dimensions essentially coincide with the edge-angle variables of  \cite{davideynard}\footnote{We are aware that although the variables
and integration measure of \cite{davideynard} are those of 2d group field theory, the integration domain for a given triangulation are different, because
of the Delaunay conditions.}. In three dimensions these data should represent the trivial holonomy conditions of the SU(2) $BF$ theory around edges of the triangulations \cite{boulatov}. In four dimensions there exist now many interesting proposals to include also the Plebanski simplicity constraints in the group field theory formalism 
\cite{BC,EPR,ELPR,LS,FK,Baratin:2011hp,Geloun:2010vj,Krajewski:2010yq,Muxin,Riello:2013bzw}.

Renormalizing group field theory might be interesting
in this respect \cite{Rivasseau:2011xg}. Indeed the fundamental aspect of the renormalization group (RG) flow is to
average over ultraviolet fluctuations, potentially leading to an effective theory. It seems our best chance to model the way in which 
the combinatorial picture of quantum gravity could morph into the more familiar continuous picture. Along the road 
general diffeomorphism invariance should be gauge-fixed. 

Any RG analysis relies on a key technical point, the introduction of \emph{scales}. 
An RG scale is defined by an ultraviolet and infrared cutoff with a fixed ratio. Although the RG flow is 
ultimately universal, different cutoffs break different invariances of the theory and lead to
very different technical aspects of the computations. The combinatorial QFT tensor picture is scale invariant: 
all degrees of freedom are treated symmetrically. Introducing scales
should break this invariance. What could be the most natural way to do that?

A conservative proposal is to break scale invariance of a combinatorial QFT at the level of the propagator only, and in a minimal way. 
This is what happens in ordinary QFT, in which local interactions have no scale. Scales are most easily defined by dividing the 
propagator spectrum through a partition of unity.
Furthermore the Laplace (or Dirac) operators that complement the mass term in the quadratic part of an ordinary QFT action 
break locality in a minimal way. Indeed they are the ones 
which can be represented in e.g. a lattice  
discretization by \emph{nearest neighbor couplings}\footnote{Remark that for the same reason
inverse Laplacians are also the most convergent propagators in the ultraviolet compatible with 
Osterwalder-Schrader positivity, the key property which relates Euclidean to Minkovski QFT.
Indeed the Markovian character of the heat kernel responsible for the OS property is a direct consequence of the nearest neighbor
representation of the lattice Laplacian.}.

Hence to minimally break scale invariance in  the tensorial world we are also lead to the search for
a `background independent" analog of the Laplacian (or Dirac) operator to supplement
the canonical mass term so that the corresponding propagator 
\emph{propagates}. This is easily done in group field theory since Lie groups have canonical Laplacians\footnote{Lie groups also have an interesting 
$g \to g^{-1}$ symmetry, which may be a precursor for future time and parity transformations.}.
A random tensor field with values in the rank $D$ tensorial product of copies of $L^2 (G)$ spaces for a 
given Lie group $G$\footnote{Such $L^2(G)$
spaces can be truncated to a finite number of Fourier modes if we want to stick strictly to finite dimensional tensors.} can be considered also as a function on 
the space $G^D$. This point of view blurs tensorial invariance but brings in 
a canonical Laplacian for such a field. To incorporate such an operator in the QFT action, although not standard practice in group field theory, 
has been proposed from several different points of views \cite{Oriti:2007vf,Geloun:2011cy}.

Combining a Laplacian-based propagator with tensorial invariance of interactions leads to a new formalism, 
called tensorial group field theory (TGFT) \cite{Carrozza:2012uv}. The Feynman amplitudes of TGFT's are based on the 
combinatorial Feynman graphs of the random tensor models, but they include also integrals over the Lie group factors. 
The role of the Laplacian, which breaks exact tensorial invariance, is to launch
the RG flow which should connect the combinatorial and continuous pictures. The combinatorial theory could be the best description 
in the extreme ultraviolet quantum gravity regime of the theory 
(which it is tempting to identify physically with transplanckian scales). The traditional continuum description of quantum gravity
involving coordinates on a continuous manifold could be a kind of 
infrared hydrodynamic approximation. The TGFT formalism might be the best way
to understand the flow in between these two pictures.

In this direction recent research was characterized by the vigorous development of super-renormalizable and just renormalizable TGFTs.
We have now a preliminary classification of such models \cite{Geloun:2013saa,Samary:2012bw}, which  include in particular just renormalizable
models in four dimensions \cite{BenGeloun:2011rc,Geloun:2012fq} and a just renormalizable tensorial version of the SU(2) three dimensional Boulatov model \cite{Carrozza:2013wda}. It was also established that these just renormalizable tensor models are generically asymptotically free \cite{BenGeloun:2012pu,BenGeloun:2012yk,Geloun:2012qn,Samary:2013xla}.

Everything we said until now only applies to Euclidean quantum gravity. But one should also address the critical
question of the emergence of time, which plays a key role in causal dynamical triangulations (see \cite{Ambjorn:2013apa} for a recent reference).
On this question there has been little progress, except for the hope \cite{Rivasseau:2012yp} to introduce in the tensor world
an analog of the Osterwalder-Schrader positivity property, which might be a kind of seed for the future emergence of time\footnote{Of course we are
fully aware that tensorial indices in tensor models and Lie group variables in TGFTs  should not be identified naively 
with the future space-time coordinates.}. 
Let us also mention that extending group field theory 
to Lorentzian metric is non-trivial but does not seem a major problem (see \cite{Krajewski:2012aw}
for a review which includes a discussion of the Lorentzian EPRL-FK model, and \cite{Riello:2013bzw} 
for computation of radiative correction of the elementary melon graph in Lorentzian signature). This question of time is clearly one of the most pressing issue to study in tensor group field theories.

This paper is structured as follows. The next section contains a brief review of the $1/N$ expansion for vector and matrix models and
of the results on the exact counting and GH limits of the leading graphs for these two expansions. The results pertaining to the tensor track
itself are then summarized in two sections. The first one deals with tensor models and is divided into five subsections: generalities, combinatorics and polynomials, 
constructive aspects, single and double scaling limits, and statistical mechanics applications. 
The second one deals with the tensorial group field theories (TGFT), and is divided in four sections: TGFTs without gauge projector, TGFTs with gauge projector, phase transitions, and Osterwalder-Schrader positivity.
The conclusion lists many open questions.

\section{Vector and Matrix Models}
\subsection{Random Tree and Vector Models}

The number $C_n $ of rooted plane trees with $n$ edges satisfies the recursion $C_{n+1} = \sum_{i=0}^n  C_i C_{n-i}$ (obtained by cutting the left-most branch of the rooted tree)
with starting point $C_0 =1$. This identifies $C_n$ with the Catalan sequence\footnote{Early discovered by the Chinese mathematician Antu Ming in XVIIIth century.} 
$C_n  = \frac{1}{n+1}   \begin{pmatrix} 2n\\ n \end{pmatrix} $: 1, 2, 5, 14, 42...
The corresponding generating function can be computed exactly as
\bee
T(z) = \sum_n C_n z^n =  \frac{1}{{2 z}}  [1 - \sqrt{1 - 4 z}] .
\label{catalt}
\ee

The probabilistic study of rooted plane trees, both as growth processes (Galton-Watson trees) and at 
critical equilibrium, culminated in the work of Aldous \cite{aldous} who proved that the equi-distributed limit 
of such plane trees converges in the Gromov-Hausdorff sense to a universal random space, now called the 
continuous random tree or CRT.

\begin{figure}[h]
\centerline{\includegraphics[width=8 cm]{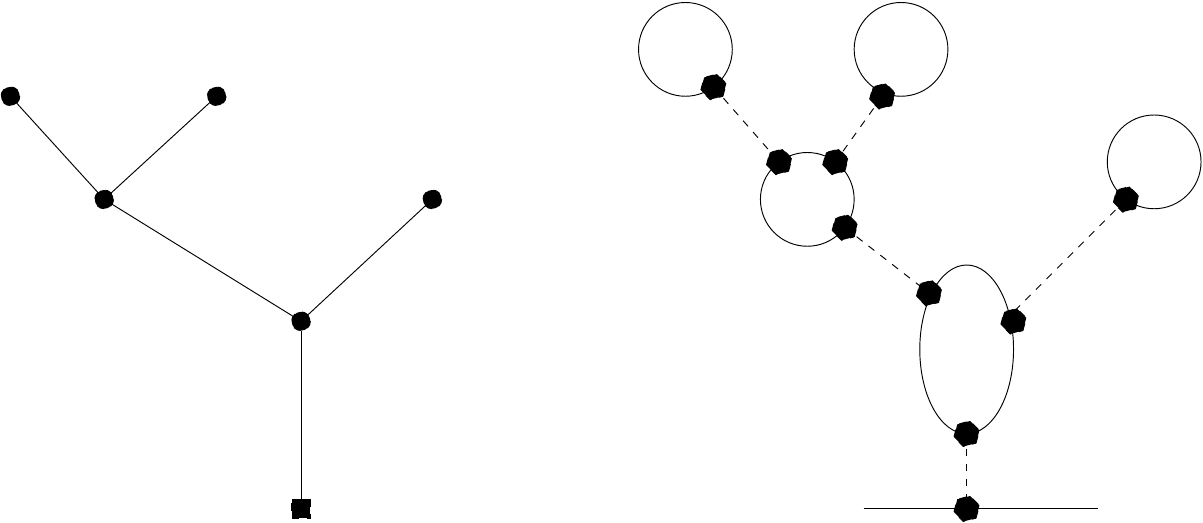}}
\caption{A rooted plane tree and the corresponding vector model graph (in the intermediate field representation).}
\end{figure}

Rooted plane trees also govern the $1/N$ expansion of $N$-vector models. 
More precisely, consider a pair of conjugate vector fields $\{\phi_i \}, \{\bar \phi_i \},  i=1 ,\cdots , N $, 
with $\lambda (\bar \phi \cdot \phi )^2$ interaction. The corresponding functional integral
\bee  Z(\lambda, N) = \frac{1}{(2i\pi )^N} \int e^{- (\bar \phi \cdot \phi )  - \lambda (\bar \phi \cdot \phi )^2 } d \bar \phi d \phi
\ee
rewrites, using a scalar intermediate field $\sigma$, as:
\bea \label{spliscalar}  Z(\lambda, N) &=&  \frac{1}{(2i\pi )^N}  \int d \sigma \frac{e^{- \sigma^2 /2 } }{\sqrt{2\pi}} \int
 e^{- (\bar \phi \cdot \phi ) + i \sqrt {2\lambda} (\bar \phi \cdot \phi ) \sigma } 
d \bar \phi d \phi\\
&=& \int  \frac{d \sigma }{\sqrt{2\pi}} e^{- \sigma^2 /2 - N \log (1 - i \sqrt {2\lambda}   \sigma )} . \nonumber
\eea
Rescaling $\lambda \to \lambda/N$ and $\sigma \to \tau = \sigma / \sqrt N $, and defining $z = -2 \lambda $ one gets
\bea  Z(\lambda, N) &=&  \int \frac{d \sigma }{\sqrt{2\pi}} e^{- \sigma^2 /2 - N \log (1 - i \sqrt {2\lambda/N}   \sigma )} \nonumber \\
&=& \sqrt N \int \frac{d \tau }{\sqrt{2\pi}}e^{- N [\tau^2 /2 + \log (1 - \sqrt {z}   \tau )]} .
\eea
A simple saddle point computation evaluates this integral as
$\frac{K e^{- N f(\tau_c) } }{ \sqrt{  f" (\tau_c)}}$,
where the saddle point of $f(\tau) = \tau^2 /2 + \log (1 - \sqrt {z}   \tau )]$ is at $\tau_c$ with $f'( \tau_c)=0$ hence 
$\tau_c = \frac{1}{{2 \sqrt z}}  [1 - \sqrt{1 - 4 z}]$. The second derivative is $f"(\tau_c) =  1 - z  (1 - \sqrt z \tau_c)^{-2}    $.

We recognize $\tau_c (z)=  \sqrt {z} T(z)$ where $T(z)$ is defined in \eqref{catalt}. More precisely
\bea \label{limpart} \lim_{N \to \infty}  \frac{ \log  Z(\lambda, N)}{N} &=&  - f(\tau_c)  = - \tau_c^2 /2 -   \log (1 - \sqrt {z} \tau_c )] \nonumber \\ 
&=&    \frac{-1}{4z} [ 1 -   \sqrt{1 - 4 z}   -2z]  - \log [ \frac{1}{2} (1 + \sqrt{1 - 4z}) ]. \nonumber
\eea
This function starts as $z/2 = - \lambda$ as expected.


As usual, the two point function is simpler (providing a root for the tree):
\bea   G^{N \to \infty}_2  
=  \frac{1}{2z}   [ 1 -   \sqrt{1 - 4 z} ]   = T(z) .
\eea

Remark also the interesting change of sign between $\lambda$ and $z$. 
$\lambda >0$ corresponds to a stable $\phi^4$ interaction (hence alternating sums), 
whether $z >0$ corresponds to  ordinary sums. $(1/N)^p$ sub-leading terms correspond to add exactly $p$ loops to the tree. 

\begin{figure}[h]
\centerline{\includegraphics[width=10cm,height=4cm,angle=0]{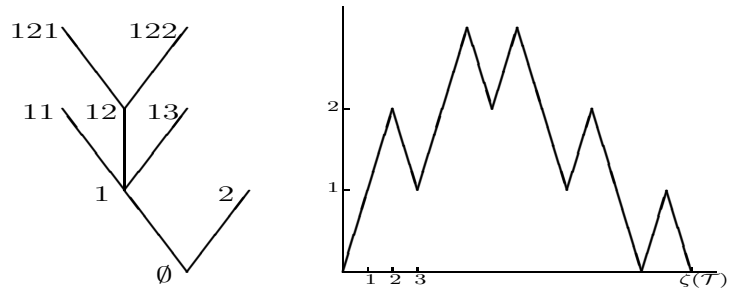}}
\caption{Labeling a tree and its height function}
\end{figure}

The metric properties of the Gromov-Hausdorff limit of equi-distributed plane trees when their number of vertices
tends to infinity are neither easy to read from the Catalan counting nor from the vector-field theoretic representation at large $N$. 
But they follow from a nice  one-to-one map between rooted plane trees and Brownian excursions.
Using the Dyck walk to turn around the tree, the tree can indeed be identified
with its height function graph (quotiented by an equivalence relation gluing the inside of the graph).
The height function is a Brownian excursion, and this is the key to prove that the Hausdorff dimension
of the CRT is $d_{Hausdorff}  = 2$. Further work proves $d_{spectral} = 4/3$ \cite{Dur}.

\subsection{Brownian Sphere and Matrix Models}

Let $Q_n$ be the number of rooted planar quadrangulations with $n$ faces.
\begin{figure}[h]
{\centerline{\includegraphics[width=7cm,angle=0]{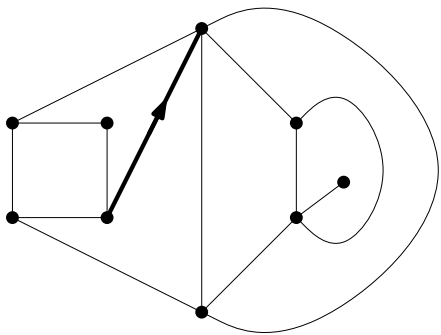}}}
\caption{A Rooted Planar Quadrangulation}
\end{figure}

Adding boundaries, Tutte found in 1962 \cite{Tutte} a quadratic recursive equation (\`a la Polchinski),
and solved it, getting the exact result:
$$Q_n = 3^n  \frac{2}{n+2}  \frac{1}{n+1}   \begin{pmatrix} 2n\\ n \end{pmatrix} . $$

It was later related to the simplest interacting complex matrix model. Consider again a pair of $N$ by $N$
complex conjugate Wishart matrices with the simplest possible interaction
\bea
Z &=&\int d M d\bar M\, \exp (-\frac12 {\rm Tr} M ^t\bar M - \frac{\lambda}{N}  {\rm Tr} M ^t\bar M M ^t\bar M  )
\nonumber   \\
&=& \sum_{n, g} a_{n,g}  (- \lambda)^n N^{2-2g}  \nonumber
\eea
where $g$ is the genus of the surface dual to the Feynman graphs in $a_{n,g}$, 
defined by the equation $ 2-2g  =  V -L +F  =  -V+F , \;\; {\rm since} \;\; L=2V. $
Planar graphs lead therefore the perturbative expansion of matrix models as $N \to \infty $ \cite{Hooft,BIPZ}.

The single scaling limit \cite{mm,Kazakov:1985ds} consists in sending $N \to \infty$-limit at fixed $\lambda$. Only planar graphs survive, and one finds 
\bee  G_{2,planar} (\lambda)  = \frac{-1 - 36 \lambda  + (1 + 24 \lambda)^{3/2} }{216 \lambda^2}  ,
\nonumber\ee
which has a singularity (critical point) at $\lambda_c  = -1/24$, with critical susceptibility $\gamma =-1/2$. 

The double scaling limit \cite{BKaz,DS,GM} sends $N \to \infty$ and $\lambda \to \lambda_c = -1/24$ simultaneous, keeping fixed $\kappa^{-1} = N^5/4(\lambda -\lambda_c )$
\bee  G_{2,double\; scaling} (\lambda)  = \sum_h  a_h \kappa^{2h} .
\ee
This sum includes all Feynman graphs, hence is not a convergent series at $\lambda >0$.
However the partition function in this double scaling obeys a Painlev\'e integrable equation, leading to an explicit 
solution called a tau function; from there on was developed the topological 2d quantum gravity theory \cite{witten,kontsevich}. 


The metric properties of the Brownian sphere can be again understood through a one-to-one map, the Schaeffer map \cite{schaeffer},
which identifies rooted \emph{pointed} planar quadrangulations with
\emph{well-labeled} oriented rooted plane trees, that is with trees equipped with a label which can either increase by one, 
decrease by one or remain fixed along each tree edge.

\begin{figure}[h]
{\centerline{\includegraphics[width=12cm,angle=0]{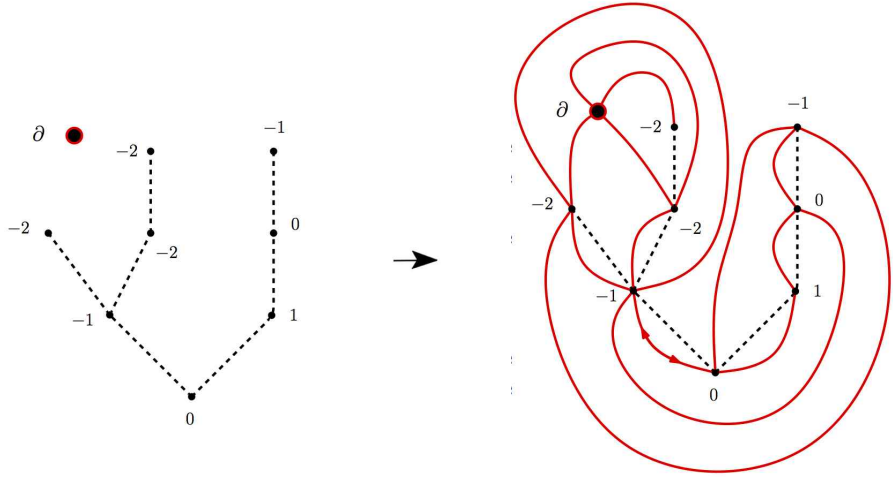}}}
\caption{The Schaeffer Map}
\end{figure}



Using the probabilistic interpretation of the Schaeffer map as a Brownian snake, 
Le Gall and Miermont proved \cite{LeGall,LeGallMiermont,Miermont} that equidistributed planar quadrangulations 
of order $n$ converge in the Gromov-Hausdorff sense as $n\to \infty$
(after rescaling  the graph distance by $n^{-1/4}$)
towards a universal random compact space, called the two-dimensional Brownian sphere.
This space has Hausdorff dimension 4 and is almost surely homeomorphic to the two-dimensional sphere.
This last result is non-trivial, as it roughly requires to prove that geodesics in typical planar graphs are not too 
close to those of the underlying tree, so that pinched spheres become of zero measure in the continuum limit.
The Brownian sphere is expected to have spectral dimension 2, although this is not proved.


To summarize, one of the simplest non trivial random geometry is the CRT or branched polymers geometry.
It has Hausdorff dimension 2, spectral dimension 4/3, and corresponds to the $1/N$ limit of vector models. A more sophisticated 
random geometry is the Brownian sphere. It has Hausdorff dimension 4, very probably spectral dimension 2, and
corresponds to the $1/N$ limit of matrix models. It is best thought of as a CRT equipped with
Schaeffer labels, which generate shortcuts on the CRT.


It is very natural to try to generalize these nice results to higher dimensions.
But many difficulties immediately arise. At the continuous level, it
is difficult to classify all geometries in dimension 3 and impossible to 
classify all (smooth) geometries in dimension 4. This is reflected at the discretized level in the fact that
it is difficult to decide whether a general triangulation in 3D is homeomorphic to the sphere $S_3$
and impossible (through a single algorithm) to decide whether a general 
triangulation in 4D is homeomorphic to the sphere $S_4$.

There are no known equivalent of the exact Catalan or Tutte counting, nor of the height function or Schaeffer map in higher dimensions. For $d \ge 3$.
we do not even know whether the number of  triangulations of the $d$-sphere with $n$ $d$-simplices is exponentially
bounded in $n$ \cite{Gromov}. There are only limited results: bounds on locally constructible triangulations \cite{DJ,PZ,ColletEckmann},
or the fact that spheres are \emph{rare} \cite{Rivasseau:2013bma}. Progress may come from unexpected directions, such as the one
of tensor models, since they do not ponder triangulations solely according to their topological class.

\section{Tensor Models}

\subsection{Generalities}

Tensor models come in two closely related basic versions: the colored model, which is essentially unique \cite{color,Gurau:2011xp}, 
and the many uncolored models \cite{Gurau:2011kk,Bonzom:2012hw}.

The colored version is made of $D+1$ tensor fields, labeled by a color index $c= 0, \cdots , D$. There are two kinds of vertices, one for fields and the other for anti-fields.
Each vertex is of degree $D+1$ and contains all colors exactly once.
Hence the Feynman graphs of this theory are bipartite $D+1$-regular edge-colored graphs. They are dual to colored $D$-dimensional triangulations.
Remark that for $D \ge 3$, such colored triangulations are better behaved than general triangulations. In particular they triangulate 
orientable\footnote{There exist also real versions of the theory
which correspond to non-orientable triangulations.} pseudo-manifolds with well defined $D$-homology \cite{Lins,FerriGagliardi,PietriPetronio,Gurau:2010nd}.
This is because the colors allow to define $p$-bubbles, for $0 \le p \le D+1$, as the connected components made of edges of the graph with $p$ distinct colors.
Vertices are connected components without any edge, $1$-bubbles are edges, $2$-bubbles are faces, which must therefore be bicolor and of even length 
and so on. It is natural to ask whether colored triangulations are not too restricted. The short answer is no. Indeed any ordinary triangulation
uniquely defines a $D$ dimensional colored triangulation, its barycentric subdivision. In particular any orientable manifold 
admits infinitely many colored triangulations\footnote{See \cite{Smerlak:2011rd,Gurau:2011cu} for interesting related discussions.}. 

The many different uncolored models correspond to effectively integrate all tensors of the colored version except one, to which we give by convention color 0,
and then truncate to a polynomial with independent coefficients (coupling constants) 
the effective interaction obtained for this last color (which is not polynomial). The polynomial interactions obtained in this way are the
vacuum graphs of the $D-1$ colored theory, hence bipartite $D$-regular edge-colored graphs.  They correspond exactly to classical $U(N)^{\otimes D}$-invariants which are 
polynomial in the tensor coefficients \cite{Collins1,Collins2}\footnote{The $U(N)^{\otimes D}$ invariance becomes 
$U(N_1) \otimes \cdots \otimes U(N_D)$ for different dimensions of the vector spaces associated to tensor indices. For simplicity the symmetric case 
$N_1 = \cdots = N_D = N$ is usually considered. Tensor invariance can be considered an
abstract "pregeometric" relativity principle: interactions should not depend on the basis in which they are written.}. 
The Feynman graphs of the theory are made of edges joining vertices which are such invariants. Expanding the
vertices as bipartite $D$-regular edge-colored graphs allows to recover $(D+1)$-regular edge-colored graphs as in the colored theory. 
Hence uncolored models with a specific finite set of interactions have a Feynman expansion which is a restriction
of the colored models expansion to graphs with a specified restricted set of vertices, i.e. $D$-bubbles of colors $1, \cdots, D$, but with independent coupling constants
for each type of vertex. 
The large $N$ limit of uncolored models with melonic interactions is Gaussian, the covariance being the full 2-point function \cite{Gurau:2011kk}, hence non-trivial.
In the case of a Bosonic statistics of the field, such uncolored models equipped with positive interactions can be defined 
non-perturbatively by constructive techniques and the previous statement can be made fully rigorous at least for 
quartic interactions \cite{Gurau:2011kk,Gurau:2013pca}. Colored models can make sense non perturbatively if all the tensor fields anticommute (or all of them save one). Such Fermionic colored models were 
the ones envisioned in the initial paper \cite{color}, which noticed the presence in that case of an interesting extra $U(D+1)$ symmetry.

\emph{Jackets} \cite{Krajewski:2010yq,Gur3,GurRiv} are ribbon graphs which encode part of the information about a
$D$-colored graph. A jacket is best defined as a cycle $\tau$ of the $D+1$ colors modulo the orientation, hence
there are  $\frac{1}{2}D!$ different jackets in a rank-$D$ tensor graph. In particular 
there are 3 jackets in $D=3$ (whose duals quadrangulate surfaces) and 12 jackets in $D=4$ (whose duals pentagulate surfaces).
The ribbon graph corresponding to a given jacket has faces with consecutive colors in the cycle. Jackets
correspond also geometrically to canonical Heegaard splittings associated to the underlying colored-triangulated space \cite{RyanSurfaces}.

The \emph{degree} of a colored graph $\cG$ is the sum of the genera of its jackets:
\bee \omega(\cG)= \sum_{\cJ} g_{\cJ}.
\ee 

After fixing the correct rescaling of the vertex, tensor models admit a $1/N$ expansion governed by the degree of the graphs. 
Indeed each face of a tensorial graph $\cG$ gives a factor $N$, each jacket contains all the vertices and all the edges of $\cG$, 
and a given face belongs to exactly $(D-1)!$ jackets, which leads to a factor $N^{- c \omega (\cG)}$ where $c$ is some constant. 
The leading order of the tensorial $1/N$ expansion (both in the colored and uncolored versions)
is given by graphs of degree $0$, called melons \cite{Bonzom:2011zz}. They triangulate the D-dimensional sphere $S_D$.
But most triangulations of the sphere $S_D$ are not melonic.

As already emphasized, the degree is \emph{not} a topological invariant of the triangulated manifold dual to the Feynman graph. The riddle that power counting,
in the related spin foam and GFT formalisms, must combine topological and cell structure information 
is neatly disentangled in the series of papers \cite{FreiGurOriti,Geloun:2010nw,BS1,BS2,BS3,Smerlak:2012je}.

Recently interesting variants of the tensorial $1/N$ expansion have been devised. In particular it was found that even rank tensors with
particular non-melonic interactions of a meandric type can lead to non-Gaussian large $N$ limits,
hence lie out of the universality class of melonic models \cite{Bonzom:2012wa}. This is linked to  meander theory \cite{Bonzom:2013lda}.

Another road to generalized models which still admit a $1/N$ expansion is given
by so-called multi-orientable models. Multi-orientable GFT has been introduced in \cite{Tanasa:2011ur} as a
simplification of three dimensional GFT which retains a larger class of Feynman graphs than colored GFT.
For the moment it can be defined only in dimension $d=3$. A further simplification (suppressing the GFT projector)
is studied in \cite{Dartois:2013he}. The corresponding rank-3 tensor model  admits a 1/N expansion in which the leading sector is still given, as in the case of colored models, 
by the melon graphs, but which differs at next-to leading order from the colored expansion.

Since tensor models have a well-defined action, their correlation functions
obey Schwinger-Dyson (SD) equations, which generalize those of random matrix theory. In the random matrix case 
these SD equations are known to form half of the Witt (Virasoro) algebra, namely its positive part. They have become standard 
tools to understand and solve matrix models, and have been fruitfully generalized beyond matrix models in the topological recursion program \cite{Borot:2013lpa}. 
In \cite{Gurau:2011tj} the SD equations for random tensor models 
have been derived in the large $N$-limit for what would soon become the uncolored version of tensor models, in which all colored fields have been integrated out save one.
The associated algebra of constraints satisfied  by the partition function were analyzed at leading order in $1/N$. The Lie algebra of these constraints 
is indexed by $D$-ary trees, hence much more complicated than in the matrix case. 
The full algebra of constraints at any order in $1/N$ was later derived
in \cite{Gurau:2012ix}. It is indexed by D-colored graphs, in which the leading order D-ary trees form a Lie subalgebra.
The constraints uniquely determine the physical perturbative solution at large $N$, as was shown in
\cite{Bonzom:2012cu}, leading to a new proof of the Gaussianity of the large $N$ limit of tensor models with melonic interactions.
Furthermore it was shown in \cite{Krajewski:2012dq} that this Lie algebra is associated to a Hopf algebra of the Connes-Kreimer type,
which means that the combinatoric rules for the insertion or contraction of tensor subgraphs within tensor graphs are well suited for RG analysis. 
The explicit construction and multi-scale analysis of renormalizable tensor field theories models (see next section) fully confirms this fact.

It is now an important goal to understand these algebras and in particular their unitary representations. It might lead
to interesting generalizations of conformal field theory in dimension higher than 2. This may be necessary in order
to correctly introduce matter fields on the spaces corresponding to the continuum limit of tensor models and find 
how their critical properties are modified in the presence of quantum gravity.

\subsection{Combinatorics, Polynomials}

Tensorial Feynman graphs are rich combinatoric objects, with more structure than ribbon graphs.
Coloring is the key property which nevertheless makes possible to somewhat tame  
their combinatorial structure.

One of the most natural question is to count tensorial invariants, in particular the connected ones.
Connected melonic graphs are in one to one correspondence with
edge-colored D-ary tree, hence can be counted by Narayana numbers \cite{Bonzom:2012zf}.
In \cite{Geloun:2013kta} a new algebraic more abstract
encoding of the tensor graphs into conjugacy classes of permutations is introduced. Allowing to use standard combinatoric tools
such as Burnside's lemma, it leads to systematic enumerative results. 
In particular the counting of tensor invariants has been related to the counting of branched covers of the 2-sphere, hence to topological field theory. The
rank $D$ of the tensor becomes the order of the  branch points. The counting of color-symmetrized tensor invariants was also obtained, as 
well as formulas for correlators of the tensor model invariants. 

Other important tools associated to Feynman graphs (introduced to solve their XIXth century precursors, namely electric circuits)
are their associated Kirchoff polynomials. In the field theory context they are called Symanzik polynomials
and allow to write the parametric representation of Feynman integrals. They are also a multivariate version of the Tutte polynomial
which is the universal solution of linear deletion/contraction recursions. In the matrix case, analog polynomials 
which satisfy simple deletion/contraction recursion have been developed for ribbon graphs by Bollob\'as, Riordan
and other authors \cite{BollobasRiordan,Gurau:2006yc,Krajewski:2008fa,Krajewski:2010pt}. 
To extend these graph polynomials to tensor models is a subject in active development.

The first attempt to define such polynomials for colored graphs is in \cite{Gurau:2009tz}. Since the category of colored graphs is not closed under the operation of contraction, this attempt is not fully satisfying. In \cite{Tanasa:2010me} another polynomial, satisfying a deletion/contraction relation, was defined, both for colorable and non-colorable graphs, which is different of the one defined in \cite{Gurau:2009tz}. 

To find a better analog of contraction in the context of colored graphs, an enlarged category of ``weakly-colored" graphs is introduced in \cite{AGH1}.
These graphs are obtained by gluing stranded vertices with stranded edges according to definite rules which generalize those of colored graphs. 
In the rank-3 case, a polynomial invariant in seven variables which obeys a contraction/cut recursion (the cut replacing the deletion operation) is 
defined for this category. It is a non-trivial generalization going beyond the Tutte or the Bollob\'as-Riordan polynomials defined now on graphs representing 3D simplicial pseudo-manifold with boundary. Some progresses towards the universal property of this polynomial has been discussed in \cite{AGH2}. 
However the explicit computation of such polynomials remains very difficult, even for simple graphs. In \cite{AGL} the Bollob\'as-Riordan polynomial is explicitly computed for a special class of ribbon graphs of the `petal-flower" type. But for instance, the set of boundary conditions
of the recurrence relation obeyed by the Bollob\'as-Riordan polynomial remains an open issue \cite{AGL}. It is expected that, for the higher rank
case, the situation becomes much more involved.

\subsection{Constructive Tensor Theory}

Constructive field theory is a set of techniques to perform the ultraviolet limit and renormalization
of a quantum field theory in a rigorous way. It allows to check that the Schwinger functions of 
a theory, typically in superrenormalizable or asymptotically free cases, are well-defined and obey all the 
Osterwalder-Schrader axioms of Euclidean quantum field theory, ensuring the existence of an 
analytic continuation to Minkovski space theory which satisfies Wightman axioms.

Historically the key tool for constructive theory has been the use of cluster/Mayer expansions,
especially in their powerful multiscale version \cite{GJ,Riv}. However cluster and Mayer expansions require to
decompose space-time into a regular lattice, such as a cubic lattice. They are therefore not well adapted to the case
of curved space-time, and fail completely for non commutative QFTs, GFTs and combinatorial QFTs,
in which the interactions are not local in any ordinary sense. 

A new method, the loop vertex expansion (LVE), was introduced to bypass this difficulty and extend constructive methods to matrix models in \cite{Rivasseau:2007fr}. 
It combines three ingredients which do not depend on any space-time background: the intermediate field method which decomposes any interaction in terms of elementary three body interactions, a symmetric forest formula \cite{BK,AR} and the use of replicas. The result recasts Bosonic field theory with stable (positive) interactions into a convergent expansion which automatically computes the Borel sum of the ordinary perturbative expansion.

Initially designed to treat the $\phi^4$ interaction (see \cite{Rivasseau:2009pi} for a pedagogical introduction), 
the LVE has been extended to stable scalar interactions in \cite{Rivasseau:2010ke}. It has been shown compatible with 
direct space decay estimates \cite{Magnen:2007uy} and with renormalization in simple superrenormalizable cases 
\cite{Rivasseau:2011df,ZWGW}.

The combinatoric core of the LVE is so simple that it can be summarized in a few lines \cite{Rivasseau:2013ova}.
One defines a set of positive weights $w(G,T)$ for any pair made of a connected graph $G$ and a spanning tree
$T \subset G$. They are normalized so as to form a probability measure on the spanning trees of $G$:
\be  \sum_{T \subset G}  w (G,T) =1  \label{bary0} .
\ee
To compute constructively instead of perturbatively a connected QFT correlation $S$ which is a sum over
connect graphs of Feynman amplitudes $A_G$ one uses \eqref{bary0} 
to introduce a sum over trees for each graph, and then simply
\emph{exchange} the order of summation between graphs and their spanning trees
\be  S = \sum_{G} A_G = \sum_{G }  \sum_{T \subset G}  w(G,T)  A_G =  \sum_{T} A_T, \quad A_T = \sum_{G \supset T}  w(G,T)  A_G .
\label{const}
\ee
The constructive `miracle" is that if one uses the \emph{right graphs} and \emph{right weights},
then for a stable Bosonic QFT with cutoffs one gets
\be  \sum_{T}  \vert A_T\vert < +\infty ,
\label{const1}
\ee
which means that $S$ is now well defined; furthermore the result is the Borel sum of the ordinary perturbation expansion.
In some sense the commutation of graphs and trees redeems the initial sin of perturbative QFT, namely the illegal exchange
of a series and a functional integral. From a physical perspective it is not too surprising after all that trees are the more 
fundamental structures and that they index better the perturbative series. The loops of Feynman graphs, after all, 
always correspond to virtual effects which are never observed: only tree-like jets of particles can be observed in collision experiments.

It remains to say which are the graphs and weights to use for \eqref{const1} to hold. The graphs are not the ones of the ordinary perturbatve series, but the ones
of the intermediate field formalism. The weights must be \emph{constructive}, i.e. obey a certain positivity property \cite{Rivasseau:2013tpa}. 
Naive equidistributed weights $w(G,T) = 1/\chi(G)$, (where $\chi (G)$ is the \emph{complexity}, i.e. number of spanning trees of $G$) 
are not constructive, but the following weights are \cite{Rivasseau:2013ova}:
\be w(G,T) =\frac{N(G,T)}{|E|!}  \label{def}
\ee
where  $N(G,T)$ is the number of Hepp sectors (complete ordering of the edges of $G$) in which $T$ is leading in the sense of Kruskal's greedy algorithm \cite{Kruskal}.

In another recent development it was shown that the weights of \eqref{def} are not the only constructive ones, but just the most symmetric
under permutation of vertices. There is a family of constructive weights
for each non-trivial partition of the set of vertices of $G$ \cite{Rivasseau:2013tpa}. The regular weights are the most symmetric, corresponding to the partition into singletons.
Other partitions may be interesting to consider, for instance in the case of the computation of connected functions.

The constructive study of tensor models started in \cite{sefu1}, where the LVE combined with inductive Cauchy-Schwarz estimates 
proved that one can Borel sum group field theory with a stable Freidel-Louapre interaction \cite{FreidelLouapre} uniformly in a domain which scales in 
the right way as $N \to \infty$. The first application to \emph{uncolored} tensor models is in \cite{Gurau:2011kk}, where the LVE was used to prove
that the interacting tensor model with the simplest quartic melonic interactions indeed has a Gaussian limit at large $N$.
Then to our surprise it was found in \cite{Gurau:2013pca} that the 
LVE, once rewritten in a kind of parametric representation,
is even better adapted to tensor than to matrix models. Indeed (in the same case of a tensor model 
with the simplest quartic melonic interactions) the leading term of the LVE, in which all intermediate fields are set to 0, computes exactly the leading
term of the $1/N$ expansion hence the melonic graphs \cite{Gurau:2013pca}. This is very different from the matrix
case in which the leading term of the LVE gives estimates of the right order of magnitude but does not compute the 
full planar series.

\begin{figure}[h] \label{cardioid}
\centerline{\includegraphics[width=8cm,angle=0]{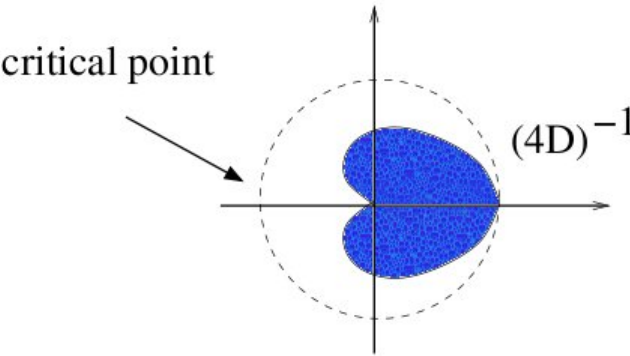}}
\caption{The cardioid domain}
\end{figure}

To summarize the difference between perturbative and constructive analysis for the two-point function of the quartically interacting uncolored model, 
one could compare the Feynman expansion, which states
\bee S_2 = 1 -D\lambda - \frac{1}{N^{D-2}} D\lambda  + \sum_G A^G(N), \quad A^G(N)  \sim \lambda^2   \label{quartic1}
\ee
to the $1/N$ expansion, which states that
\bee \label{quartic2}
S_2 =  \frac{ (1+4D\lambda )^{\frac{1}{2} } - 1 }{ 2D\lambda }   +  \sum_GA^G(N),  \qquad A^G(N) \sim \frac{1}{N^{ D-2} },
\ee
and finally to the constructive analysis which \emph{proves} that this two point function is analytic in $\lambda = |\lambda|e^{\imath \varphi}$ in the cardioid domain 
$  \vert \lambda \vert <(  \cos\frac{\varphi}{2} )^2 /8D$ of Figure \ref{cardioid}
where it can be expanded as
\bee \label{quartic3} S_2 =  \frac{ (1+4D\lambda )^{\frac{1}{2} } - 1 }{ 2D\lambda }  + \dots  +  {\cal R}^{(p)}_N(\lambda) ,
\ee 
and where the remainder ${\cal R}^{(p)}_N(\lambda)$ obeys a rigorous bound, namely 
\bee  
| {\cal R}^{(p)}_N(\lambda) | \le  \frac{ 1 }{ N^{p(D-2)} }   \frac{|\lambda|^p}{ \Bigl( \cos\frac{\varphi}{2}\Bigr)^{2p+2} } \; p! \; A \; B^p 
\ee
for some constants $A$ and $B$.

\subsection{Single and Double Scaling}

The single scaling limit of tensor models is defined, as for matrix models, by keeping only the leading graphs, namely the melons,
summing the corresponding restricted series, and searching for the closest singularity or critical point. 
The Schwinger-Dyson equations of the colored tensor model in this approximation
becomes an extremely  simple algebraic equation from which the explicit solution can be read easily. The result is again simplest for the two-point function, 
$G_{2,melo}$ restricted to melons graphs
\bea 
G_{2,melo}  = \frac{ (1+4D\lambda )^{\frac{1}{2} }- 1 }{ 2D\lambda }  ,
\eea 
where $\lambda $, the coupling constant, is the product of the coupling of the vertex and of the antivertex \cite{Bonzom:2011zz}.
This two-point function becomes critical for $\lambda\to \lambda_c =-(4D)^{-1}$. The corresponding susceptibility $\gamma=1/2$
corresponds to branched polymers. At the critical point the average number of vertices diverges, which means a continuous geometry is reached. 
Using the work of Albenque and Mackert \cite{Albenque} on Appolinian triangulations, it has been proved in \cite{Gurau:2013cbh}  that this continuous geometry is indeed a CRT (branched polymers),
with Hausdorff dimension 2 and spectral dimension 4/3. 
This is a non-trivial result, since the graph distance is the one induced by the \emph{dual triangulation}. It does not correspond to the ordinary graph distance on the Feynman graph, but rather to a 
colored analog, in which the distance counts the shortest number of full sequences containing all colors between two points \cite{Gurau:2013cbh}.

As mentioned earlier, the structure of subleading terms in the $1/N$ tensor expansion is richer than for matrices and requires careful study. This study was started
in \cite{Kaminski:2013maa} where the next-to-leading order contribution in the $1/N$ expansion was analyzed. It was shown that it has the same radius of convergence than the
leading order melonic series and a new value of the susceptibility exponent, $\gamma= 3/2$, compatible with the existence of a double scaling limit, which consists in sending both $N \to \infty$ and $\lambda \to \lambda_c$. If the rate is well chosen, 
this double scaling will select the sub-leading graphs with optimal ratio between powers of $1/N$ and powers of $\lambda - \lambda_c$. This analysis was extended
to multi-orientable models in \cite{Raasakka:2013eda}.

Tensorial double scaling was fully performed for the color model in \cite{GurauSchaeffer} and for
the uncolored model with quartic interactions in \cite{Dartois:2013sra}. In the case of the color model the notion of schemes is introduced in \cite{GurauSchaeffer}. They are the terminal forms of colored graphs under 
certain reduction algorithms which in particular trim the melonic parts. Using this notion it is established that the series of graphs which share a given fixed scheme has the same radius of convergence than the melonic series. 
Simultaneously the double scaling has also been performed in \cite{Dartois:2013sra} but for the uncolored models, with a different reduction technique; the leading family of graphs in this case
was dubbed `cherries": they are made of tree-like graphs decorated with small loops at their leaves \cite{Dartois:2013sra}.

The conclusions of both approaches agree. The double scaling in tensor models is totally different from the one of matrix models. It sums only a small
class of graphs, still exponentially bounded, hence it is stable. Moreover, and quite unexpectedly, it exhibits a qualitatively different behavior below and above dimension 6.
Below dimension 6 it displays a singularity at fixed distance from the origin and is clearly the first step in a richer set of multi-scaling limits which remains to be explored.

To summarize the double scaling results for the same quartically interacting model with coupling constant $\lambda$ that is considered
in \eqref{quartic1}-\eqref{quartic3}, we introduce the variable 
\bee
x=N^{D-2} \bigl[ (4D)^{-1} + \lambda \bigr] \Rightarrow \lambda = - \frac{1}{4D} + \frac{x}{N^{D-2} }, 
\ee
and send $N\to \infty$ and $\lambda\to - \frac{1}{4D}$ while keeping $x$ fixed. We obtain a power series in $x$
\bea
G_2 =  N^{1-\frac{D}{2} }
\sum_{p\ge 0}  \frac{c_p}{ x^{p-\frac{1}{2} } } + Rest, \qquad Rest  < N^{1/2-D/2} 
\eea 
which has a new critical point  in $x$ is at $x_c = 1 /4(D-1)$. The corresponding double scaling-limit, including melons and cherry trees 
is then
\bea \bar G_{2,cherry} (N) &=&  2 - 4  N^{(1-D/2)} \sqrt {D(x - x_c)}  + O( N^{\frac{1}{2}-\frac{D}{2}}) ,
\eea
so that the singularity remains, for $D<6$ of the branched polymer type, but at a different location \cite{Dartois:2013sra}. 
As explained above this seems a good starting point to attempt a triple scaling.

For tensor models equipped with a gauge projector of the Boulatov-Ooguri  type, much less is known. The leading graphs of the $1/N$ expansion are still melons, but no algebraic
closed form equation is known for the two-point function. For the moment the best we know is that the series satisfy both a 
lower and an upper exponential bound in the number of vertices so that there must be a phase transition
\cite{Baratin:2013rja}. These bounds were obtained in terms of a graph polynomial related to a higher dimensional generalization of the Kirchhoff tree-matrix theorem. 

\subsection{Statistical Mechanics Applications}\label{statmech}

Like random matrices, random tensors are so general that they should have rich applications, 
not only to quantum gravity but also to many other areas of physics and mathematics. The first such domain of applications is certainly
statistical mechanics, since random tensor graphs can be considered as random lattices which can support added
statistical variables.

The first attempt in this direction \cite{Bonzom:2011ev} studied the Ising model coupled to random tensor geometry in the footsteps of Kazakov \cite{kazakovising}.
The idea was to generalize the two-matrix model for Ising spins on random surfaces into a two-tensor model. 
In agreement with numerical investigations, in the continuum limit no phase transition was found. This is consistent with
the idea that the CRT space cannot support a phase transition
for an Ising model without long range couplings.

To force the phase transition it is possible to modify the model. In \cite{Benedetti:2011nn} a new class of colored tensor models was introduced with a modified 
propagator associating weight factors to the faces of the graphs, where curvature is concentrated. 
The leading order in $1/N$ has been solved analytically, and it was shown that a particular model
undergoes a third-order phase transition in the continuum
characterized by a jump in the susceptibility exponent.

More generally multi-critical points can be studied, which generalize the multi-critical behavior of matrix models.
They can be interpreted as models of dimers on a set of random
lattices for the sphere in dimension three and higher. Dimers with their exclusion rules are generated
by the different interactions between tensors, whose coupling constants are dimer activities. 
In \cite{Bonzom:2012np} a particular example of a multi-critical point is interpreted as a transition between the
dilute phase and a crystallized phase, though with negative activities.
In \cite{Bonzom:2012qx} hard dimers on dynamical lattices in arbitrary dimensions are studied again as a random tensor model,
with exact calculation of their critical exponents, and an alternative description as a system
of Òcolor-sensitive hard-core dimersÓ on random branched polymers is provided.

Perhaps one of the most promising direction, although little developed yet, is to apply
the random tensor models to the theory of spin glasses. Here one of the main goal is to go from
the "mean-field" infinite dimensional Sherrington-Kirckpatrick model towards a more realistic model
showing finite dimensional behavior such as the Edwards-Anderson model \cite{spinglass}. In this respect the 
$p$-spin model introduced in \cite{GrossMezard} uses independent identically distributed
interactions between $p$ different spins for $p \ge 3$, hence is exactly a free tensor model.
In \cite{BGSspinglass} the case of an interacting tensor model for spin glass is considered.
Adapting the results on the Gaussian character of the large $N$ limit of these models 
it can be proved that such `$p$-spin models with non-Gaussian quenched interactions" 
belong to the same universality class than the free $p$-spin model. Moreover the change of critical temperature
due to the interactions can be studied precisely. This seems
one of the very few rigorous result of this kind in this domain. It is tempting to 
speculate that going beyond the leading $1/N$ behavior, for instance through multiple scaling 
limits, might have to do with interesting effects.

Another interesting connection has been developed recently between random tensors and
meander enumeration. A meander is a self-avoiding path crossing a river through an even number of $2n$
bridges. Meander enumeration is connected to matrix models, hence two-dimensional gravity \cite{DiFrancesco:1999tp}.
But it is also connected to rank-4 tensor models with the particular invariants first considered in \cite{Bonzom:2012wa}, in which two sets of two colors 
have a single face. Such invariants are labeled by permutations. Considering them as
interactions and expanding the corresponding tensor theory into Feynman graphs, 
one obtains a problem equivalent to counting meanders whose lower arch configuration is obtained from the upper arch configuration by a permutation on half of the arch feet.
It has been possible to reduce this computation to those of interactions labeled by stabilized-interval-free permutations,
which correspond  to the enumeration of so-called \emph{irreducible} meandric systems. 

\section{Tensorial Group Field Theories}

\subsection{Laplacian Propagator}

The key difference between random matrix models and NCQFTÕs lies in the modification of the propagator.
Breaking the U(N) invariance of matrix models at the propagator level leads to renormalizable non-commutative quantum field theories such as the Grosse-Wulkenhaar 
(GW) model \cite{Grosse:2004yu} on ${\mathbb R}^4$ equipped with the Moyal product. It is a matrix model with quartic coupling and a propagator which can be interpreted
as the sum of the Laplacian and an harmonic potential on ${\mathbb R}^4$, but which is very natural in the matrix base which expresses the Moyal star product as a matrix product (see however \cite{Gurau:2008vd} for an alternative possibility).
The GW propagator breaks the U(N) invariance of the theory in the infrared (small N) regime, but is asymptotically 
invariant in the ultraviolet regime (large N); hence it launches an RG flow between these two regimes. Power counting in such models is entirely governed by the underlying $1/N$ expansion (divergent graphs are the 
regular planar graphs), hence the corresponding matrix renormalization group \cite{Eichhorn:2013isa} can be considered a kind of continuous version of the $1/N$ expansion. It is remarkable that this matrix RG flow is 
asymptotically safe \cite{Grosse:2004by,Disertori:2006uy,Disertori:2006nq}. 
Its planar sector has also recently been beautifully solved \cite{Grosse:2009pa,Grosse:2012uv}, showing that four-dimensional integrability, like four dimensional asymptotic safety, is not limited 
to the famous example of supersymmetric Yang-Mills $N=4$ QFT but is more general and does not \emph{require} supersymmetry.

In a completely analogous manner, one can define TGFT's with tensorial interactions and a soft breaking of the tensorial invariance of their propagator. 
Renormalization is again a continuous version of the $1/N$ expansion and the divergent graphs are the melons. 
The group structure allows to interpret the fields both as rank $D$ tensors and as fields defined on $G^D$ where $G$ is the Lie group. The propagator is the inverse of the sum of the Laplacians
on each factor in $G^D$. In the simplest case, namely $G=U(1)$, one can identify at any rank which are the super-renormalizable and just renormalizable interactions. 
Surprisingly perhaps, at rank/dimension 4 the just renormalizable interactions are of order 6, not 4 \cite{BenGeloun:2011rc,Geloun:2012fq,BenGeloun:2012pu}. The quartically interacting model is just 
renormalizable at rank/dimension 5. It is also just renormalizable at
rank/dimension 3 if the propagator has Dirac rather than Laplace power counting.
It is remarkable that the corresponding tensorial RG flow generically displays the physical property of asymptotic freedom 
in all renormalizable known cases \cite{BenGeloun:2012pu,BenGeloun:2012yk,Geloun:2012qn}.

The bridge between matrix and tensor renormalizable models has been further reduced in \cite{Geloun:2012bz},
in which new families of renormalizable  models are identified. It is shown that the rank 3 tensor model defined in 
\cite{BenGeloun:2012pu}, and the Grosse-Wulkenhaar models in two and four dimensions generate three different
classes of renormalizable models by modifying the power of the propagator. 
A recent review on this class of models extends and generalizes this analysis \cite{Geloun:2013saa}.
It classifies the models of matrix or tensor type with a propagator which is the inverse of a sum of momenta of the form $p^{2a}$, $a \in ]0,1]$, 
hence exactly those for which the Osterwalder-Schrader positivity can hold. 
Infinite towers of (super- and just-) renormalizable matrix models are found. 
Moreover, so far, all renormalizable tensor models are proved to be asymptotically free and several matrix models are shown asymptotically safe at one-loop and, very likely, at any loop (some of them possess however a Landau ghost). The question of further restricting the space of models is still under current investigation. 

The emerging picture is a simple classification of quite a large class of Bosonic renormalizable non-local QFTs of the matrix or tensor type, which was completely unexpected a few years ago.  The theories of scalar and vector type are generically not asymptotically free since they have
no wave function renormalization at one loop. The theories of matrix type are generically asymptotically safe (specially for instance all $Tr[M^4]$ models), 
since the wave function renormalization exactly compensates the coupling constant renormalization. The theories of tensor type are generically asymptotically free since the wave function renormalization wins over
the coupling constant renormalization.

The combinatorics of renormalization is neatly encoded in a Connes-Kreimer combinatorial Hopf algebra. See \cite{Krajewski:2012is} for 
a version of this algebra adapted to multi-scale analysis.
This algebra has been explicited in \cite{Raasakka:2013kaa} for the case of the model defined in 
\cite{BenGeloun:2011rc,Geloun:2012fq}. It is significantly different from previous Connes-Kreimer algebras,
due to the involved combinatorial and topological properties of the tensorial Feynman graphs.

\subsection{Laplacian with Gauge Projector}

TGFTs which include a ``Boulatov-type" projector in their propagator \cite{boulatov} deserve a category of their own as they are both more difficult to renormalize and closer to the desired geometry in the continuum. Their characteristic feature is to include an average over a new Lie group variable acting simultaneously over all tensor threads in the propagator. In a Feynman graph, once the vertex variables have been integrated out, the Feynman amplitude appears as an integral over one variable for each edge
of a product of delta functions for each face. Each such delta function fixes to the identity the ordered product of the edge variables around the face.
Therefore it can be interpreted as a theory of simplicial geometry supplemented with a discrete gauge connection at the level 
of the Feynman amplitudes of the theory which has trivial holonomy
around each face of the Feynman graph, hence around each $D-2$ cell of the dual triangulation.

To introduce renormalizable models of this class is again done in two steps.
First one replaces the usual interaction of the Boulatov model by tensor invariant interactions to allow for a simple power-counting of the $1/N$ type.
Second, one adds to the usual GFT Gaussian measure a Laplacian term to allow for scale analysis.
It is however truly non-trivial that renormalization can still work for models of this type. Indeed their
propagator $C( g_1, \cdots, g_D; g'_1, \cdots, g'_D ) $ is definitely \emph{not} asymptotic 
in the ultraviolet regime to a product of delta functions
$\delta(g_1 (g'_1)^{-1})  \cdots \delta(g_D (g'_D)^{-1})$ but rather to an average:
\bee C( g_1,\cdots, g_D; g'_1, \cdots, g'_D )  \simeq_{uv} \int dh \; \delta(g_1h (g'_1)^{-1})  \cdots \delta(g_Dh (g'_D)^{-1}).
\ee
Nevertheless, and to the author's own surprise, an extension of the locality principle still holds for divergent graphs,
meaning that they can be renormalized again by tensor 
invariant counterterms. This 
is possible because the divergent graphs to renormalize, which are melonic graphs, precisely  have a sufficiently particular structure for such 
an extended locality property to hold, although it does not hold at all for general graphs. 
It seems that tensor theory at each step provides an unexpected cure for its problems.

The set of results  include the study of a family of Abelian TGFT models in 4d which was shown to be super-renormalizable for any polynomial interaction 
\cite{Carrozza:2012uv}, so is similar in power counting to the family of $P(\phi)_2$ models in ordinary QFT. Abelian just renormalizable models also exist 
in 5 and 6 dimensions and were studied in \cite{Samary:2012bw}. They are asymptotically free \cite{Samary:2013xla}.
Finally an even more interesting non-Abelian model with group manifold $SU(2)$ was successfully renormalized in \cite{Carrozza:2013wda} and is
also expected to be asymptotically free. For an excellent review of this subject we refer to \cite{Carrozza:2013mna}.

Let us recall that the above studies concern \emph{uncolored} TGFTs. 
For the colored Boulatov model, Ward-Takahashi identities were derived  and interpreted in
\cite{BenGeloun:2011xu}. In \cite{Geloun:2013zka} the ultraviolet behavior of the same colored theory
but with a Laplacian added to the propagator is studied. It is shown that  all orders in perturbation theory in the case of the $U(1)$ group 
in three dimensions are convergent; moreover it was convincingly argued that this finiteness should also hold  for the same model over $SU(2)$.  
Recall however that colored Bosonic models are a priori non-perturbatively unstable, and that colored Fermionic models are more natural.

\subsection{Phase Transitions and Cosmology}

Renormalizability of TGFT's is not a goal \emph{per se}. It is a tool to study associated phase transitions, which hopefully
should become at some point relevant for physics and in particular for cosmology. 

A theory is considered physical if the associated computations agree with experiments. But to deserve being called a theory
it needs also to be robust in some RG sense.
Particular cutoffs, particular discretizations, particular truncations, particular numerical experiments only capture part of the physical content 
of a theory; ultimately all these schemes are related through RG universality. 
For instance lattice QCD, although it is up to now the best way to extract numerical values for glueballs or hadron masses, is not considered the full physical
theory of QCD. We do not believe that there exist little lattices inside protons. 
We believe that quarks exist and that QCD is a physical theory of strong interactions because all its 
different cutoffs and expansion schemes up to now agree between themselves and with several different experiments, 
and this is what RG universality can explain.

Hence just as asymptotic ultraviolet freedom of QCD has its infrared counterpart in quark confinement and nuclear physics,
we hope the RG flow of TGFTs (and in particular of the renormalizable asymptotically free TGFTs) to lead to a physically interesting cosmology.
Suppose an observer could run down the RG flow of such a fundamental theory. 
He would see the universe, initially\footnote{"Initially" in the sense of scales; there can be RG scales "before" there is time \cite{Rivasseau:2011hm}.} 
in its "pregeometric" phase, condense through the big-bang into an effective continuous space-time and cool with all its matter fields through many phase transitions 
and accretions to the present universe. Conversely, running the movie backwards, that is against the RG flow,
he would see the universe dissolve into more and more elementary constituents 
and ultimately see space-time itself ``boil" into a pregeometric phase described by the fundamental 
theory, in our case the perturbative regime of a TFGT. 

In \cite{Konopka:2008hp} the condensation of a pregeometric universe into the one we live in 
was dubbed ``geometrogenesis" and the name graphity was coined 
to suggest that it could come from a purely combinatorial approach on graphs. 

In group field theory this scenario of space-time emerging as a condensate 
has been extensively developed by D. Oriti and collaborators.
In \cite{Oriti:2013jga} a general discussion is given on the 
disappearance of continuum space and time at microscopic scale. The paper outlines the related
conceptual issues and shows how the GFT framework for quantum gravity can provide a concrete 
computational approach to geometrogenesis.

In \cite{Gielen:2013kla}, this approach is further elaborated, in order to extract the 
dynamics of the condensate states directly from the 
GFT dynamics, following the same procedure used in ordinary quantum fluids. 
The effective dynamics provides a nonlinear and nonlocal 
extension of quantum cosmology. A GFT model with a kinetic term of Laplacian type is shown to give rise, 
in a semiclassical (WKB) approximation and in the isotropic case, to a modified 
Friedmann equation. Clearly this sets up a promising exploration program for the future. It
could only benefit from the improved TGFT formalism, which brings the full panoply of QFT tools, 
including renormalization and constructive theory, to further study the geometrogenesis scenario.
 
\subsection{OS Positivity}

In \cite{Rivasseau:2012yp} it was proposed that Osterwalder-Schrader (OS) positivity has an analog in the tensor world. 
Little progress has been made so far on this idea. As we know, in ordinary QFT OS positivity is related to 
analytic continuation to real time, unitarity and causality \cite{OS}; in 
a purely statistical mechanics context it is also the source of checkerboard estimates \cite{GRS}. However 
it does not seems \emph{a priori} compatible with non-local interactions. Let us discuss informally why.

A general ordinary Euclidean QFT is a functional measure over fields of the form $d\nu = e^{-S_{int}}   d\mu_C$, where $d\mu_C$ is the Gaussian part, $C$ being the covariance or propagator, and $S_{int}$ the interaction part. Obviously any measure of this form defines an associated positive scalar product on functions of the  field,
namely $<F.G> = \int d\nu  F \bar  G $. But only some measures have a less obvious positivity, called OS positivity. It involves a mirror symmetry.
Suppose the space on which the fields are defined admits a decomposition into two half-spaces $H_+ \cup H_-$, with a bijection $x\to Rx$
between them and a frontier $H$ that any continuous path from $H_+$ to $H_-$ must cross. In the usual case of ${\mathbb R}^d$, $H$ is any hyperplane.
OS positivity means that the bilinear form   $<F.G>_R = \int d\nu F   R(\bar G ) $ should also be a positive scalar product between functions $F$ and $G$ of fields \emph{restricted} to the half-space $H_+$. It is the existence of \emph{this} mirror scalar product which is critical to allow the continuation of such theories to Minkovski space, 
and the existence of a unitary time evolution in ordinary QFT. In short it is the Euclidean seed for the analytic continuation to real Lorentzian time\footnote{This
real-time continuation also requires the decay of Euclidean Schwinger functions when their arguments are separated, ensuring
the uniqueness of the thermodynamic limit. Indeed one needs to insert not only a separating hyperplane but also a thick separating \emph{band} of width $t$ to define the Euclidean semi-group $e^{-{\cal H}t}$, where ${\cal H} \ge 0$ 
is the Hamiltonian. It then leads to the unitary time evolution $e^{i{\cal H}t}$ in Minkovski space. We do not discuss this more technical aspect here.}.

Formally Euclidean quantum field theories are OS positive if two conditions are met, namely Markovian property for the
propagator $C$ and locality of the interaction $S_{int}$. The Markovian property ensures that the free Gaussian measure itself
is OS positive. Because any functional of the field on the half space $H_+$ can be approximated by its values at finitely many points,
OS positivity of this free Gaussian measure is equivalent to positivity of the matrix $C(x_i, Rx_j)$ for any finite set of points $x_1, ... x_n$ in $H_+$. 
Remark that positivity of the matrix $C(x_i, x_j)$ without the reflection operator $R$ means positivity of the covariance $C$ itself, hence in the 
usual translation invariant case, it just means the point-wise positivity of its Fourier transform $\hat C (p)$. This simpler positivity  is true for
any `covariance with cutoff" such as $(-\Delta + m^2)^{-k}$, where $\Delta$ is the Laplacian. For any polynomial local interaction and any dimension,
we could always find $k$ large enough so that the corresponding Euclidean measure exists and has no ultraviolet divergencies. The beauty
of QFT comes from the OS positivity condition which forbids such cutoff theories; true QFTs have therefore ultraviolet divergencies because
they must be OS positive. The covariance $(-\Delta + m^2)^{-1}$ is the best, most ultraviolet convergent propagator, with this OS positivity. Why?

This is because $C = (-\Delta + m^2)^{-1}$ is the most ultraviolet convergent propagator with the Markov property. Indeed we have the heat kernel or random walk representation
\bee
C (x,y) = \int_0^\infty dt \; \Gamma_t,  \quad \Gamma_t = \int_{\omega: x \to y } dP_t (\omega)
\ee 
which states that $C$ is an integral over times $t$ of the heat kernel $\Gamma_t$ which is itself a sum over random paths $\omega$ going from $x$ to $y$ in time $t$.
This is also why the OS positivity is most conveniently seen in a lattice discretization of the underlying space: here the random path representation
is also discretized into lattice random walks and the Markovian character is most clearly seen since lattice Green's functions for the Laplacian
can be represented in terms of random walks made of  \emph{independent nearest neighbor jumps}.

More precisely the Markov property for the Green's function of the Laplacian can be written in terms of heat kernels with Dirichlet 
boundary conditions and a first and a last ``hitting time" at which the random path $\omega$ hits the hyperplane $H$. If this hyperplane
separates the regions $H_+$ and $H_-$ in Euclidean space, and 
$x$ and $y$ are two points in $H_+$, any path $\omega$ from $x \in H_+$ to $Ry\in H_-$ admits a first hitting time $t_1$
and a last hitting time $t_1 +t_2$, so that
\bee  \Gamma_t(x,Ry) = \int_{t_1 + t_2 + t_3 =t} 
\int_{H \times H} dz_1 dz_2 \Gamma^{D,+}_{t_1} (x, z_1) \Gamma_{t_2} (z_1, z_2) \Gamma^{D,-}_{t_3} (z_2, Ry) ,
\ee
where $\Gamma^{D,+}_{t}$ is the heat kernel on $H_+$ with Dirichlet boundary conditions on $H$,
$\Gamma^{D,-}_{t}$ is the heat kernel on  $H_-$ with Dirichlet boundary conditions on $H$, 
and $\Gamma_t$ is the full free heat kernel.
Using the reflection symmetry and Markovian property of the heat kernel, we have
\bee  \Gamma_t(x,Ry) = \int_{t_1 + t_2 + t_3 =t} 
\int_{H \times H} dz_1 dz_2 \Gamma^{D,+}_{t_1} (x, z_1) \Gamma_{t_2} (z_1, z_2)  \Gamma^{D,+}_{t_3} (z_2, y) ,
\ee
where the last heat kernel has been reflected. This formula, now symmetric between $x$ and $y$, means that
the matrix $C(x_i, Rx_j)$ for any finite set of points $x_1, ... x_n$ in $H_+$ is indeed positive as announced.

If we turn to the measure $d\nu$, because the interaction is local, it factorizes over $H_+$ and $H_-$ as
\bee\label{local}
e^{-S_{int}}   = e^{-S_{int}^+} \;\;  e^{-S_{int}^-} .
\ee  

OS positivity can then be interpreted as a decomposition of the functional measure. Considering $d\mu^H$
the free measure on "boundary" fields $\psi$ living on $H$ with covariance $C (z_1, z_2)$, we obtain the formal gluing formula
\bee \int F_+ (\phi) \bar G_+ (R\phi)  d\nu (\phi) =  \int d\mu^H  (\psi ) \int_{} F_+ (\phi) \bar G_+ (\phi) d\nu_+ (\phi) ,
\ee
where $d\nu_+ ( \phi)$ is the functional integral on fields supported on $H_+$ with interaction $e^{-S_{int}^+}$, $\psi$ is the boundary value of $\phi$ 
and $d\mu^H$ corresponds to the measure  $d \nu $ but restricted to fields $\psi$  living on the separating hyperplane $H$.
Hence the scalar product $<F G>_R$ is an integral of positive scalar products, hence is positive.

In practice the difficult part of constructive field theory is to prove existence of the measure $d\nu$ by a limit process. Typically
different types of cutoffs violate different OS axioms; in the ultraviolet limit all of these axioms hold because this limit is usually unique. 
A nice simple way to prove this unicity is to prove that the limit is the Borel sum of the perturbative theory.

We understand now why OS positivity looks very problematic for QFT's with non-local interactions: equation \eqref{local} simply no longer holds.
Nevertheless we think that for uncolored tensor models and 
TGFTs with positive \emph{melonic} invariant interactions OS positivity can be extended in a natural way.
The first step is easy. A Lie group admits a reflection $g \to g^{-1}$. On fields defined in $G^D$, the free Gaussian measure with
covariance $C$ inverse of sums of group Laplacians as in \cite{BenGeloun:2011rc}, should inherit OS positivity from the 
Markovian character of the Laplacian heat kernel\footnote{As usual for a lattice OS positivity, this property should hold only for the hyperplanes perpendicular to
the lattice axis, hence here for reflections $g \to g^{-1}$ on each of the tensor field coordinates.}.
Certainly the non-locality of tensorial interactions seems a harder problem but here also a solution is in sight: 
melonic interactions can probably be inductively decomposed using intermediate fields of the matrix type, 
as in \cite{Gurau:2013pca} so that one may for a given axis always find a decomposition of the interaction factoring into the left and right side of the hyperplane.
At least for the simplest quartically interacting model considered in \cite{Gurau:2013pca}, OS positivity should follow \cite{DGR}. It would be also
very interesting  to know whether this OS positivity resists the imposition of a Boulatov-type gauge projection on the propagator. We recall again
that OS positivity in tensor space should not be confused with ordinary OS positivity on space-time. It is a new concept which provides a non 
trivial scalar product on tensorial observables and we propose to explore its consequences.

\section{Conclusion: Open Questions}

The list of open questions on the tensor track agenda is growing fast. Here are a few of them formulated in quite general terms, 
perhaps reflecting our personal interests and bias. 

\begin{itemize}

\item  Develop the algebraic formulation of tensor invariants in terms of permutations
as started in \cite{Geloun:2013kta}. Classify and count the positive (i.e. mirror symmetric) invariants.

\item Understand the joint distribution of eigenvalues for the intermediate field matrices of \cite{Gurau:2013pca}.

\item Investigate the multiple scaling limits of colored and uncolored models.

\item Adapt to the TGFT context the many tools used to study phase transitions in QFT.

\item Study the symmetry algebra of the uncolored tensor models and their representation theory. 
Can one define a smaller closed algebra associated to positive (stable) interactions?

\item Connect rank-D tensor models to D-brane theory. How do extended melonic objects 
behave in a background geometry?

\item Both the vertices and external observables of rank $D$ tensor models are rank $(D-1)$ vacuum tensor graphs. Does this connect to the 
holographic principle in quantum gravity?

\item Do renormalizable models of the EPRL-FK type exist in four dimensions and under which conditions?

\item Define uncolored models with higher than quartic interactions, prove Borel-Le Roy 
summability of their perturbative series.

\item Extend the cardioid domain of Figure \ref{cardioid} so as to obtain the cut along the negative axis. 

\item Develop the OS positivity condition. Is it a useful axiom for tensor theory? Can it be related to
unitarity and to a real time formalism?

\item Develop the statistical mechanics applications of section \ref{statmech}.

\end{itemize}

Through work on these questions we hope the tensor track can also serve as a way to better connect 
the many different approaches to quantum gravity.

\medskip\noindent
{\bf Acknowledgments}

\medskip
This review has benefitted from fruitful discussions with many people, in particular M. Albenque, A. Baratin, J. Ben Geloun, V. Bonzom, S. Carrozza, N. Curien, S. Dartois, B. Dittrich, F. David, B. Duplantier, B. Eynard, L. Freidel, H. Grosse, R. Gurau, T. Krajewski, J.F. Le Gall, R. Loll, D. Oriti, G. Schaeffer, M. Raasakka, A. Riello, C. Rovelli, J. Ryan, D.O. Samary, M. Smerlak, A. Tanasa, F. Vignes-Tourneret and R. Wulkenhaar.
I thank especially J. Ben Geloun, V. Bonzom, S. Dartois, R. Gurau and D. Oriti for a critical reading, and I also thank the organizers of the Corfu School that led to this review.



\begin{thebibliography}{99}

\bibitem{Rivasseau:2011hm}
V.~Rivasseau,
``Quantum Gravity and Renormalization: The Tensor Track,''
arXiv:1112.5104.

\bibitem{Rivasseau:2012yp} 
V.~Rivasseau,
``The Tensor Track: an Update,''
arXiv:1209.5284.

\bibitem{boulatov} 
D.~V.~Boulatov,
``A Model of three-dimensional lattice gravity,''
Mod.\ Phys.\ Lett.\ A {\bf 7}, 1629 (1992),
hep-th/9202074.
 
\bibitem{laurentgft}
L.~Freidel,
``Group field theory: An overview,''
Int.\ J.\ Theor.\ Phys.\  {\bf 44}, 1769 (2005),
arXiv:hep-th/0505016.
  
\bibitem{Oriti:2011jm}
D.~Oriti,
``The microscopic dynamics of quantum space as a group field theory,''
arXiv:1110.5606.
  
\bibitem{Krajewski:2012aw} 
T.~Krajewski,
``Group field theories,''
PoS QGQGS {\bf 2011}, 005 (2011),
arXiv:1210.6257. 
  
\bibitem{Baratin:2010wi}
A.~Baratin and D.~Oriti,
``Group field theory with non-commutative metric variables,''
Phys.\ Rev.\ Lett.\  {\bf 105}, 221302 (2010), arXiv:1002.4723. 
  
\bibitem{Baratin:2011aa} 
A.~Baratin and D.~Oriti,
``Ten questions on Group Field Theory (and their tentative answers),''
J.\ Phys.\ Conf.\ Ser.\  {\bf 360}, 012002 (2012), arXiv:1112.3270.

\bibitem{Oriti:2013aqa} 
  D.~Oriti,
  ``Group field theory as the 2nd quantization of Loop Quantum Gravity,''
  arXiv:1310.7786 [gr-qc].
  
\bibitem{Rovelli}
Carlo Rovelli, ``Quantum Gravity", Cambridge University Press 
(2004), ISBN 0-521-83733-2.

\bibitem{Douglas:2001ba} 
M.~R.~Douglas and N.~A.~Nekrasov,
``Noncommutative field theory,''
Rev.\ Mod.\ Phys.\  {\bf 73}, 977 (2001),
hep-th/0106048.
  
\bibitem{Rivasseau:2007ab}
V.~Rivasseau,
``Non-commutative renormalization,''
arXiv:0705.0705.

\bibitem{Mehta}
M. L. Mehta,  ``Random Matrices,'' Elsevier, Pure and Applied Mathematics (Amsterdam), 142 (2004). 

\bibitem{Di Francesco:1993nw}
  P.~Di Francesco, P.~H.~Ginsparg and J.~Zinn-Justin,
  ``2-D Gravity and random matrices,''
  Phys.\ Rept.\  {\bf 254}, 1 (1995)
  arXiv:hep-th/9306153.
  
\bibitem{ambjorn-book}
  J.~Ambjorn, B.~Durhuus and T.~Jonsson,
  ``Quantum geometry. A statistical field theory approach,''
Cambridge University Press, (1997).

\bibitem{scratch}
R. Loll, J. Ambjorn and J. Jurkiewicz,
``The Universe from Scratch",
Contemp.Phys. 47 (2006) 103-117, hep-th/0509010,

\bibitem{Ambjorn:2013apa} 
  J.~Ambjorn, A.~G\"orlich, J.~Jurkiewicz and R.~Loll,
  ``Causal dynamical triangulations and the search for a theory of quantum gravity,''
  Int.\ J.\ Mod.\ Phys.\ D {\bf 22}, 1330019 (2013).

\bibitem{Gurau:2011kk}
  R.~Gurau,
  ``Universality for Random Tensors,''
  arXiv:1111.0519.

\bibitem{Gur3}
  R.~Gurau,
``The 1/N expansion of colored tensor models,''
  Annales Henri Poincar\'e {\bf 12}, 829 (2011),
  arXiv:1011.2726. 

\bibitem{GurRiv}
R.~Gurau and V.~Rivasseau,
``The 1/N expansion of colored tensor models in arbitrary dimension,''
  Europhys.\ Lett.\  {\bf 95}, 50004 (2011),
  arXiv:1101.4182. 

\bibitem{Gur4}
  R.~Gurau,
  ``The complete 1/N expansion of colored tensor models in arbitrary
  dimension,''
  arXiv:1102.5759. 

\bibitem{Gurau:2011xp}
  R.~Gurau and J.~P.~Ryan,
  ``Colored Tensor Models - a review,''
  arXiv:1109.4812.

\bibitem{color}
 R.~Gurau,
  ``Colored Group Field Theory,''
  Commun.\ Math.\ Phys.\  {\bf 304}, 69 (2011),
  arXiv:0907.2582. 
  
\bibitem{Geloun:2009pe} 
  J.~Ben Geloun, J.~Magnen and V.~Rivasseau,
  ``Bosonic Colored Group Field Theory,''
  Eur.\ Phys.\ J.\ C {\bf 70}, 1119 (2010),
  arXiv:0911.1719. 

\bibitem{Seiberg:2006wf}
N.~Seiberg,
``Emergent space-time,''
arXiv:hep-th/0601234.
 
\bibitem{Oriti:2006ar}
D.~Oriti, ``A quantum field theory of simplicial geometry and the emergence of space-time,''
J.\ Phys.\ Conf.\ Ser.\  {\bf 67}, 012052 (2007),
arXiv:hep-th/0612301.

\bibitem{Oriti:2007qd}
  D.~Oriti,
``Group field theory as the microscopic description of the quantum space-time
  fluid: a new perspective on the continuum in quantum gravity,''
  arXiv:0710.3276. 

\bibitem{Konopka:2008hp}
  T.~Konopka, F.~Markopoulou and S.~Severini,
 ``Quantum Graphity: a model of emergent locality,''
  Phys.\ Rev.\  D {\bf 77}, 104029 (2008),
  arXiv:0801.0861. 
  
\bibitem{Sindoni:2011ej}
  L.~Sindoni,
 ``Emergent models for gravity: an overview,''
  arXiv:1110.0686. 
 
\bibitem{gromov}
M. Gromov, ``Metric structures for Riemannian and non-Riemannian spaces", Birkh\"auser (1999).


\bibitem{causalsets}
J. Henson,  ``The causal set approach to quantum gravity", In D. Oriti,
editor, Approaches to Quantum Gravity: Towards a New Understanding
of Space and Time, Cambridge University Press, 2006.

\bibitem{aldous} D. Aldous, ``The Continuum Random Tree I, II and III", The Annals of Probability, 1991, Vol 19, 1-28; in 
Stochastic Analysis, London Math Society Lecture Notes, Cambridge University Press 1991n eds Barlow and Bingham;
The Annals of Probability, 1993, Vol 21, 248-289.
  
\bibitem{LeGall}   J.F. Le Gall, 
``Uniqueness and universality of the Brownian map",
arXiv:1105.4842.

\bibitem{LeGallMiermont} J.-F. Le Gall, G. Miermont,
``Scaling limits of random trees and planar maps",
arXiv:1101.4856.

\bibitem{Miermont}  G. Miermont,
``The Brownian map is the scaling limit of uniform random plane quadrangulations",
arXiv:1104.1606.

\bibitem{ADT1} 
J. Ambjorn, B. Durhuus and T. Jonsson, ``Summing over all genera for $d > 1$: a toy model", Phys. Lett. B 244
(1990), 403, 412.

\bibitem{ADT2} 
J. Ambjorn, B. Durhuus and T. Jonsson, ``Three-Dimensional Simplicial Quantum Gravity 
And Generalized Matrix Models,Ó Mod. Phys. Lett. A 6, 1133 (1991).

\bibitem{sasa} 
N. Sasakura, ``Tensor model for gravity and orientability of manifold,Ó Mod. Phys. Lett. A 6, 2613
(1991).

\bibitem{gross} 
M. Gross, ``Tensor models and simplicial quantum gravity in $>$ 2-D,Ó Nucl. Phys. Proc. Suppl. 25A,
144 (1992).


\bibitem{ambjorn}
J. Ambjorn, ``Simplicial Euclidean and Lorentzian Quantum Gravity", arXiv:gr-qc/0201028.

\bibitem{Gurau:2011tj}
R.~Gurau,
``A generalization of the Virasoro algebra to arbitrary dimensions,''
Nucl.\ Phys.\  B {\bf 852}, 592 (2011),
arXiv:1105.6072. 

\bibitem{Bonzom:2012hw} 
  V.~Bonzom, R.~Gurau and V.~Rivasseau,
  ``Random tensor models in the large N limit: Uncoloring the colored tensor models,''
  Phys.\ Rev.\ D {\bf 85}, 084037 (2012),
  arXiv:1202.3637. 

\bibitem{Geloun:2013kta} 
  J.~B.~Geloun and S.~Ramgoolam,
  ``Counting Tensor Model Observables and Branched Covers of the 2-Sphere,''
  arXiv:1307.6490.
  
\bibitem{Bonzom:2011zz}
  V.~Bonzom, R.~Gurau, A.~Riello and V.~Rivasseau,
 ``Critical behavior of colored tensor models in the large N limit,''
  Nucl.\ Phys.\  B {\bf 853}, 174 (2011),
  arXiv:1105.3122. 
    
\bibitem{Bonzom:2012wa} 
  V.~Bonzom,
  ``New 1/N expansions in random tensor models,''
  JHEP {\bf 1306}, 062 (2013),
  arXiv:1211.1657. 

\bibitem{Gurau:2013cbh} 
  R.~Gurau and J.~P.~Ryan,
  ``Melons are branched polymers,''
  arXiv:1302.4386. 

\bibitem{GurauSchaeffer} 
R. Gurau and G. Schaeffer,
  ``Regular colored graphs of positive degree,"
arXiv:1307.5279.
 
\bibitem{Dartois:2013sra} 
  S.~Dartois, R.~Gurau and V.~Rivasseau,
  ``Double Scaling in Tensor Models with a Quartic Interaction,''
  arXiv:1307.5281. 
  
\bibitem{Gurau:2011sk} 
  R.~Gurau,
  ``The Double Scaling Limit in Arbitrary Dimensions: A Toy Model,''
  Phys.\ Rev.\ D {\bf 84}, 124051 (2011),
  arXiv:1110.2460. 

\bibitem{Carlip:2009kf} 
  S.~Carlip,
  ``Spontaneous Dimensional Reduction in Short-Distance Quantum Gravity?,"
  arXiv:0909.3329. 
  
\bibitem{Stojkovic}
D. Stojkovic, ``Vanishing dimensions: theory and phenomenology",  arXiv:1304.6444.
    

\bibitem{Baratin:2011tg} 
  A.~Baratin, F.~Girelli and D.~Oriti,
  ``Diffeomorphisms in group field theories,''
  Phys.\ Rev.\ D {\bf 83}, 104051 (2011),
  arXiv:1101.0590. 
    
\bibitem{Polyakov1981207}
A. Polyakov, 
``Quantum geometry of bosonic strings", Physics Letters B, 103(3): 207-210 (1981).

\bibitem{Friedan:1982is}
D. Friedan, ``Introduction to Polyakov's String Theory",
in J. Zuber and R. Stora editors, Recent advances in field
theory and statistical mechanics, volume~39 of Les Houches summer
school of theoretical physics, Amsterdam. North-Holland (1984).

\bibitem{KPZ:1988}
V. Knizhnik, A. Polyakov and A. Zamolodchikov, ``Fractal structure of 2d-gravity",
Mod. Phys. Lett. A, 03:819, (1988).

\bibitem{David1988a} F. David, ``Conformal field theories coupled to 2-d gravity in the conformal gauge", Mod. Phys. Lett. A, 03:1651-1656.

\bibitem{David1988b} F. David, ``Sur l'entropie des surfaces al\'eatoires", C. R. Acad. Sci. Paris, 307 S\'erie II: 1051-1053.

\bibitem{DistlerKawai} J. Distler and H. Kawai, ``Conformal field theory and 2-d quantum gravity", Nucl. Phys. B 1989, 321:509.

\bibitem{Duplantier} B.~Duplantier,
``Conformal Random Geometry", Les Houches 2005 Lecture Notes, arXiv:math-ph/0608053.

\bibitem{Duplantier:2009np} 
B.~Duplantier and S.~Sheffield,
``Duality and KPZ in Liouville Quantum Gravity,''
  Phys.\ Rev.\ Lett.\  {\bf 102}, 150603 (2009),
  arXiv:0901.0277.
  
\bibitem{DuplantierSheffield}
B. Duplantier, S. Sheffield,
``Liouville Quantum Gravity and KPZ", Inventiones mathematicae
2011, Vol. 185, Issue 2, 333-393, arXiv:0808.1560.
  
\bibitem{davideynard}
F. David and B. Eynard, ``Planar maps, circle patterns and 2d gravity",
arXiv:1307.3123.

\bibitem{Plebanski} 
J. F. Plebanski, ``On the separation of Einsteinian substructures", Journal of Mathematical Physics, 18:2511-
2520, (1977).

\bibitem{Holst} 
S. Holst, ``BarberoÕs Hamiltonian derived from a generalized Hilbert-Palatini action", Physical Review D,
53, 5966-5969 (1996), arXiv:gr-qc/9511026.

\bibitem{Krasnov} 
K. Krasnov, ``Plebanski Formulation of General Relativity: A Practical Introduction", 2009, arXiv:0904.0423.

\bibitem{Gielen:2010cu} 
  S.~Gielen and D.~Oriti,
  ``Classical general relativity as BF-Plebanski theory with linear constraints,''
  Class.\ Quant.\ Grav.\  {\bf 27}, 185017 (2010), arXiv:1004.5371.

\bibitem{BC} 
J. W. Barrett and L. Crane. ``Relativistic spin networks and quantum gravity," Journal of Mathematical
Physics, 39:3296-3302, (1998), arXiv:gr-qc/9709028.

\bibitem{EPR}
 J. Engle, R. Pereira, and C. Rovelli, ``Loop-quantum-gravity vertex amplitude,''  Phys. Rev. Lett.  {\bf 99},
161301 (2007), arXiv:0705.2388.

\bibitem{ELPR}  J. Engle, E. Livine, R. Pereira, and C. Rovelli, ``LQG vertex with finite Immirzi parameter",
Nucl. Phys. B, 799:136-149, 2008, arXiv:0711.0146.

\bibitem{LS} E. R. Livine and S. Speziale, ``Solving the simplicity constraints for spinfoam quantum gravity," Europhysics
Letters, 81:50004, March 2008, arXiv:0708.1915.

\bibitem{FK}
 L. Freidel and K. Krasnov,  ``A new spin foam model for 4d gravity,''  Class. Quant. Grav. {\bf 25},
125018 (2008), arXiv:0708.159.

\bibitem{Baratin:2011hp} 
  A.~Baratin and D.~Oriti,
  ``Group field theory and simplicial gravity path integrals: A model for Holst-Plebanski gravity,''
  Phys.\ Rev.\ D {\bf 85}, 044003 (2012), arXiv:1111.5842.
  
\bibitem{Geloun:2010vj} 
  J.~Ben Geloun, R.~Gurau and V.~Rivasseau,
  ``EPRL/FK Group Field Theory,''
  Europhys.\ Lett.\  {\bf 92}, 60008 (2010), arXiv:1008.0354.
 
\bibitem{Krajewski:2010yq} 
  T.~Krajewski, J.~Magnen, V.~Rivasseau, A.~Tanasa and P.~Vitale,
  ``Quantum Corrections in the Group Field Theory Formulation of the EPRL/FK Models,''
  Phys.\ Rev.\ D {\bf 82}, 124069 (2010), arXiv:1007.3150.
  
\bibitem{Muxin}
Muxin Han, Mingyi Zhang,
``Asymptotics of Spinfoam Amplitude on Simplicial Manifold: Euclidean Theory",
arXiv:1109.0500.

\bibitem{Riello:2013bzw} 
  A.~Riello,
  ``Self-Energy of the Lorentzian EPRL-FK Spin Foam Model of Quantum Gravity,''
  Phys.\ Rev.\ D {\bf 88}, 024011 (2013),
  arXiv:1302.1781. 

\bibitem{Rivasseau:2011xg} 
  V.~Rivasseau,
  ``Towards Renormalizing Group Field Theory,''
  PoS CNCFG {\bf 2010}, 004 (2010),
  arXiv:1103.1900.

\bibitem{Oriti:2007vf} 
  D.~Oriti and T.~Tlas,
  ``A New Class of Group Field Theories for 1st Order Discrete Quantum Gravity,''
  Class.\ Quant.\ Grav.\  {\bf 25}, 085011 (2008),
  arXiv:0710.2679. 
  
\bibitem{Geloun:2011cy} 
  J.~Ben Geloun and V.~Bonzom,
  ``Radiative corrections in the Boulatov-Ooguri tensor model: The 2-point function,''
  Int.\ J.\ Theor.\ Phys.\  {\bf 50}, 2819 (2011),
  arXiv:1101.4294. 
  
\bibitem{Carrozza:2012uv} 
  S.~Carrozza, D.~Oriti and V.~Rivasseau,
  ``Renormalization of Tensorial Group Field Theories: Abelian U(1) Models in Four Dimensions,''
  arXiv:1207.6734. 

\bibitem{Geloun:2013saa} 
  J.~Ben Geloun,
  ``Renormalizable Models in Rank $d\geq 2$ Tensorial Group Field Theory,''
  arXiv:1306.1201. 

\bibitem{Samary:2012bw} 
  D.~O.~Samary and F.~Vignes-Tourneret,
  ``Just Renormalizable TGFT's on $U(1)^{d}$ with Gauge Invariance,''
  arXiv:1211.2618. 

\bibitem{BenGeloun:2011rc}
  J.~Ben Geloun and V.~Rivasseau,
``A Renormalizable 4-Dimensional Tensor Field Theory,''
  arXiv:1111.4997. 
  
\bibitem{Geloun:2012fq} 
  J.~Ben Geloun and V.~Rivasseau,
  ``Addendum to 'A Renormalizable 4-Dimensional Tensor Field Theory',''
  Commun.\ Math.\ Phys.\  {\bf 322}, 957 (2013),
  arXiv:1209.4606. 
  
\bibitem{Carrozza:2013wda} 
  S.~Carrozza, D.~Oriti and V.~Rivasseau,
  ``Renormalization of an SU(2) Tensorial Group Field Theory in Three Dimensions,''
  arXiv:1303.6772. 

\bibitem{BenGeloun:2012pu} 
  J.~Ben Geloun and D.~O.~Samary,
  ``3D Tensor Field Theory: Renormalization and One-loop $\beta$-functions,''
  Annales Henri Poincare {\bf 14}, 1599 (2013),
  arXiv:1201.0176. 

\bibitem{BenGeloun:2012yk} 
  J.~Ben Geloun,
  ``Two and four-loop $\beta$-functions of rank 4 renormalizable tensor field theories,''
  Class.\ Quant.\ Grav.\  {\bf 29}, 235011 (2012),
  arXiv:1205.5513. 
    
\bibitem{Geloun:2012qn} 
  J.~Ben Geloun,
  ``Asymptotic Freedom of Rank 4 Tensor Group Field Theory,''
  arXiv:1210.5490.
  
\bibitem{Samary:2013xla} 
  D.~O.~Samary,
  ``Beta functions of $U(1)^d$ gauge invariant just renormalizable tensor models,''
  arXiv:1303.7256. 

\bibitem{Dur}
B. Durhuus, T. Jonsson and J. Wheater, 
``The spectral dimension of generic trees," 
J. Stat. Phys. 128 (2007) 1237-1260. math-ph/0607020.

\bibitem{Tutte} W. T. Tutte, ``A census of planar triangulations",
Canad. J. Math. 14 (1962), 21-38.

\bibitem{Hooft}
 G. 't Hooft, ``A PLANAR DIAGRAM THEORY FOR STRONG INTERACTIONS,'' Nucl. Phys. B  {\bf 72},
 461 (1974).
 
\bibitem{BIPZ}
  E.~Brezin, C.~Itzykson, G.~Parisi and J.~B.~Zuber,
  ``Planar Diagrams,''
  Commun.\ Math.\ Phys.\  {\bf 59}, 35 (1978).

\bibitem{mm}
   F.~David,
  ``A Model of Random Surfaces with Nontrivial Critical Behavior,''
  Nucl.\ Phys.\ B {\bf 257}, 543 (1985).
  
\bibitem{Kazakov:1985ds}
  V.~A.~Kazakov,
  ``Bilocal Regularization of Models of Random Surfaces,''
  Phys.\ Lett.\ B {\bf 150}, 282 (1985).

\bibitem{BKaz}
 E.~Br\'ezin and V.~A.~Kazakov, 
``Exactly solvable field theories of closed strings,''
Phys. \ Lett.\  B {\bf 236}, 144 (1990).

\bibitem{DS}
M.~R.~Douglas and S.~H.~Shenker, 
``Strings in less than one dimension,'' 
Nucl. \ Phys. \ B {\bf 335}, 635 (1990).

\bibitem{GM}
D.~J.~Gross  and  A.~A.~Migdal, 
``Nonperturbative two-dimensional quantum gravity,'' 
Phys. \ Rev. \ Lett. {\bf 64}, 127 (1990).

\bibitem{witten}   
E. Witten, ``On the structure of the topological phase of two
dimensional gravity", Nucl. Phys. B  1990, 340:281.

\bibitem{kontsevich}  M.  Kontsevich, ``Intersection theory on the moduli space of
curves and the matrix Airy function", Commun. Math. Phys., 1992, 147:1-23.

\bibitem{schaeffer}  
G. Schaeffer, ``Conjugaison d'arbres et cartes combinatoires al\'eatoires", PhD thesis 1998.

\bibitem{Gromov} M. Gromov. 
``Spaces and questions," Geom. Funct. Anal. (2000), 118-161GAFA 2000 (Tel Aviv, 1999).

\bibitem{DJ}
B. Durhuus and T. J\'onsson, ``Remarks on the entropy of 3-manifolds," Nuclear Phys. B 445 (1995),
182-192.
  
\bibitem{PZ} 
J. Pfeiße and G. M. Ziegler, ``Many triangulated 3-spheres", Math. Ann. 330 (2004), 829-837.

\bibitem{ColletEckmann}
P Collet, J.-P. Eckmann, M. Younan,
``Trees of nuclei and bounds on the number of triangulations of the 3-ball",
arXiv:1204.6161.

\bibitem{Rivasseau:2013bma} 
  V.~Rivasseau,
``Spheres are rare,''
  arXiv:1303.7371. 

\bibitem{Lins}   
S. Lins, ``Gems, computers and attractors for 3-manifolds", Series on Knots and Everything, Vol. 5, World
Scientific Publishing Co. Inc., River Edge, NJ, 1995.

\bibitem{FerriGagliardi}  
M. Ferri and C. Gagliardi, ``Crystallization moves", Pacific J. Math. 100 (1982) 85-103. 

\bibitem{PietriPetronio}   
R. De Pietri, C. Petronio, J. Math. Phys. 41, 6671-6688 (2000), arXiv:gr-qc/0004045.

\bibitem{Gurau:2010nd} 
  R.~Gurau,
  ``Lost in Translation: Topological Singularities in Group Field Theory,''
  Class.\ Quant.\ Grav.\  {\bf 27}, 235023 (2010),
  arXiv:1006.0714. 

\bibitem{Smerlak:2011rd} 
  M.~Smerlak,
  ``Comment on `Lost in Translation: Topological Singularities in Group Field Theory',''
  Class.\ Quant.\ Grav.\  {\bf 28}, 178001 (2011),
  arXiv:1102.1844. 
 
\bibitem{Gurau:2011cu} 
  R.~Gurau,
  ``Reply to comment on `Lost in translation: topological singularities in group field theory',''
  Class.\ Quant.\ Grav.\  {\bf 28}, 178002 (2011), arXiv:1108.4966.

\bibitem{Collins1}  B. Collins, ``Moments and cumulants of polynomial random variables on unitary groups, the Itzykson- Zuber integral, and free probability,Ó Int. Math. Res. Not. 17, (2003) 953, arXiv:math-ph/0205010.

\bibitem{Collins2} 
B. Collins and P. Sniady, ``Integration with respect to the Haar measure on unitary, orthogonal and symplectic group,Ó Commun. Math. Phys. 264, 773 (2006), arXiv:math-ph/0402073.

\bibitem{Gurau:2013pca} 
  R.~Gurau,
  ``The $1/N$ Expansion of Tensor Models Beyond Perturbation Theory,''
  arXiv:1304.2666.

\bibitem{RyanSurfaces} 
J. P. Ryan, ``Tensor models and embedded Riemann surfaces", Physical Review D, 2012, arXiv:1104.5471.

\bibitem{FreiGurOriti}
  L.~Freidel, R.~Gurau and D.~Oriti,
  ``Group field theory renormalization - the 3d case: power counting of
  divergences,''
  Phys.\ Rev.\  D {\bf 80}, 044007 (2009),
  arXiv:0905.3772. 

\bibitem{Geloun:2010nw} 
  J.~Ben Geloun, T.~Krajewski, J.~Magnen and V.~Rivasseau,
  ``Linearized Group Field Theory and Power Counting Theorems,''
  Class.\ Quant.\ Grav.\  {\bf 27}, 155012 (2010),
  arXiv:1002.3592. 

 \bibitem{BS1}
  V.~Bonzom and M.~Smerlak,
 ``Bubble divergences from cellular cohomology,''
  Lett.\ Math.\ Phys.\  {\bf 93}, 295 (2010)

\bibitem{BS2}
 V.~Bonzom and M.~Smerlak,
``Bubble divergences from twisted cohomology,''
arXiv:1008.1476. 

\bibitem{BS3}
  V.~Bonzom and M.~Smerlak,
 ``Bubble divergences: sorting out topology from cell structure,''
  arXiv:1103.3961. 

\bibitem{Smerlak:2012je} 
  M.~Smerlak,
  ``Divergences in spinfoam quantum gravity,''
  arXiv:1201.4874. 

\bibitem{Bonzom:2013lda} 
  V.~Bonzom and F.~Combes,
  ``Fully packed loops on random surfaces and the 1/N expansion of tensor models,''
  arXiv:1304.4152.
  
\bibitem{Tanasa:2011ur} 
  A.~Tanasa,
  ``Multi-orientable Group Field Theory,''
  J.\ Phys.\ A {\bf 45}, 165401 (2012),
  arXiv:1109.0694.

\bibitem{Dartois:2013he} 
  S.~Dartois, V.~Rivasseau and A.~Tanasa,
  ``The 1/N expansion of multi-orientable random tensor models,''
  arXiv:1301.1535.
  
\bibitem{Borot:2013lpa} 
  G.~Borot, B.~Eynard and N.~Orantin,
  ``Abstract loop equations, topological recursion, and applications,''
  arXiv:1303.5808. 

\bibitem{Gurau:2012ix} 
  R.~Gurau,
 ``The Schwinger Dyson equations and the algebra of constraints of random tensor models at all orders,''
  Nucl.\ Phys.\ B {\bf 865}, 133 (2012),
  arXiv:1203.4965.
  
\bibitem{Bonzom:2012cu} 
  V.~Bonzom,
 ``Revisiting random tensor models at large N via the Schwinger-Dyson equations,''
  JHEP {\bf 1303}, 160 (2013),
  arXiv:1208.6216. 
     
\bibitem{Krajewski:2012dq} 
  T.~Krajewski,
 ``Schwinger-Dyson Equations in Group Field Theories of Quantum Gravity,''
  arXiv:1211.1244. 
 
\bibitem{Bonzom:2012zf} 
  V.~Bonzom and R.~Gurau,
  ``Counting Line-Colored D-ary Trees,''
  arXiv:1206.4203. 
   
\bibitem{BollobasRiordan}   
B. Bollob\'as and O. Riordan, ``Polynomial of graphs on surfaces,Ó Math. Ann. 323, 81-96 (2002).
  
\bibitem{Gurau:2006yc} 
  R.~Gurau and V.~Rivasseau,
  ``Parametric representation of noncommutative field theory,''
  Commun.\ Math.\ Phys.\  {\bf 272}, 811 (2007), math-ph/0606030.
  
\bibitem{Krajewski:2008fa} 
  T.~Krajewski, V.~Rivasseau, A.~Tanasa and Z.~Wang,
  ``Topological Graph Polynomials and Quantum Field Theory, Part I: Heat Kernel Theories,''
  J.\ Noncommut.\ Geom.\  {\bf 4}, 29 (2010), arXiv:0811.0186.
  
\bibitem{Krajewski:2010pt} 
  T.~Krajewski, V.~Rivasseau and F.~Vignes-Tourneret,
  ``Topological graph polynomials and quantum field theory. Part II. Mehler kernel theories,''
  Annales Henri Poincare {\bf 12}, 483 (2011), arXiv:0912.5438.

\bibitem{Gurau:2009tz} 
  R.~Gurau,
 ``Topological Graph Polynomials in Colored Group Field Theory,''
  Annales Henri Poincar\'e {\bf 11}, 565 (2010), arXiv:0911.1945.
  
  
\bibitem{Tanasa:2010me} 
  A.~Tanasa,
  ``Generalization of the Bollob\'as-Riordan polynomial for tensor graphs,''
  J.\ Math.\ Phys.\  {\bf 52}, 073514 (2011),
  arXiv:1012.1798 [math.CO].

\bibitem{AGH1}
R. C. Avohou, J. Ben Geloun and M. N. Hounkonnou,
 ``A Polynomial Invariant for Rank 3 Weakly-Colored Stranded Graphs",
arXiv:1301.1987.

\bibitem{AGH2}
R. C. Avohou, J. Ben Geloun and M. N. Hounkonnou,
 ``Universality for polynomial invariants on ribbon graphs with flags",
arXiv:1310.3708.
 
\bibitem{AGL}
R. C. Avohou, J. Ben Geloun and E. R. Livine 
 ``On terminal forms for topological polynomials for ribbon graphs: The $N$-petal flower",
arXiv:1212.5961.

\bibitem{GJ} 
J. Glimm and A. M. Jaffe, ``Quantum Physics. A Functional Integral Point Of View,Ó New York, Springer (1987). 

\bibitem{Riv}
V. Rivasseau, ``From perturbative to constructive renormalization,Ó
Princeton University Press (1991). 

\bibitem{Rivasseau:2007fr} 
  V.~Rivasseau,
  ``Constructive Matrix Theory,''
  JHEP {\bf 0709}, 008 (2007),
  arXiv:0706.1224. 

\bibitem{BK} 
D. Brydges and T. Kennedy, ``Mayer expansions and the Hamilton-Jacobi equation", Journal of Statistical Physics, 48, 19 (1987).

\bibitem{AR}
A. Abdesselam and V. Rivasseau, ``Trees, forests and jungles: A botanical garden for cluster expansionsÓ, in
Lecture Notes in Phys., 446, Springer, Berlin, 1995; arXiv:hep-th/9409094.

\bibitem{Rivasseau:2009pi} 
  V.~Rivasseau,
  ``Constructive Field Theory in Zero Dimension,''
  Adv.\ Math.\ Phys.\  {\bf 2010}, 180159 (2010),
  arXiv:0906.3524.

\bibitem{Rivasseau:2010ke} 
  V.~Rivasseau and Z.~Wang,
  ``Loop Vertex Expansion for Phi**2K Theory in Zero Dimension,''
  J.\ Math.\ Phys.\  {\bf 51}, 092304 (2010),
  arXiv:1003.1037.

\bibitem{Magnen:2007uy} 
  J.~Magnen and V.~Rivasseau,
  ``Constructive phi**4 field theory without tears,''
  Annales Henri Poincar\'e {\bf 9}, 403 (2008),
  arXiv:0706.2457.
  
\bibitem{Rivasseau:2011df} 
  V.~Rivasseau and Z.~Wang,
  ``Constructive Renormalization for $\Phi^{4}_2$ Theory with Loop Vertex Expansion,''
  J.\ Math.\ Phys.\  {\bf 53}, 042302 (2012),
  arXiv:1104.3443.
  
\bibitem{ZWGW}
Z.T. Wang, ``Construction of 2-dimensional Grosse-Wulkenhaar Model", arXiv:1104.3750.

\bibitem{Rivasseau:2013ova} 
  V.~Rivasseau and Z.~Wang,
  ``How to Resum Feynman Graphs,''
  arXiv:1304.5913. 

\bibitem{Rivasseau:2013tpa} 
  V.~Rivasseau and A.~Tanasa,
  ``Generalized constructive tree weights,''
  arXiv:1310.2424. 

 \bibitem{Kruskal} 
J. B. Kruskal, ``On the Shortest Spanning Subtree of a Graph and the Traveling Salesman ProblemÓ, Proceedings of the American 
Mathematical Society, 7, 4850, (1956).

\bibitem{sefu1}
  J.~Magnen, K.~Noui, V.~Rivasseau and M.~Smerlak,
  ``Scaling behavior of three-dimensional group field theory,''
  Class.\ Quant.\ Grav.\  {\bf 26}, 185012 (2009),
  arXiv:0906.5477. 
  
\bibitem{FreidelLouapre}
L. Freidel and D. Louapre, Phys. Rev. D68, 104004 (2003), hep-th/0211026.
  
\bibitem{Albenque} 
M. Albenque, J. F. Marckert, ``Some families of increasing planar maps," Electronic Journal of Probability 13(56),
(2008), 1624-1671.
  
\bibitem{Kaminski:2013maa} 
  W.~Kaminski, D.~Oriti and J.~P.~Ryan,
  ``Towards a double-scaling limit for tensor models: probing sub-dominant orders,''
  arXiv:1304.6934.

\bibitem{Raasakka:2013eda} 
  M.~Raasakka and A.~Tanasa,
  ``Next-to-leading order in the large N expansion of the multi-orientable random tensor model,''
  arXiv:1310.3132. 
  
\bibitem{Baratin:2013rja} 
  A.~Baratin, S.~Carrozza, D.~Oriti, J.~P.~Ryan and M.~Smerlak,
  ``Melonic phase transition in group field theory,''
  arXiv:1307.5026.

\bibitem{Bonzom:2011ev} 
  V.~Bonzom, R.~Gurau and V.~Rivasseau,
  ``The Ising Model on Random Lattices in Arbitrary Dimensions,''
  Phys.\ Lett.\ B {\bf 711}, 88 (2012),
  arXiv:1108.6269. 

\bibitem{kazakovising}
V. A. Kazakov, ``Ising model on a dynamical planar random lattice: Exact solution,Ó Phys. Lett. A
119, 140 (1986).
  
\bibitem{Benedetti:2011nn} 
  D.~Benedetti and R.~Gurau,
  ``Phase Transition in Dually Weighted Colored Tensor Models,''
  Nucl.\ Phys.\ B {\bf 855}, 420 (2012),
  arXiv:1108.5389. 
  
\bibitem{Bonzom:2012np} 
  V.~Bonzom,
  ``Multicritical tensor models and hard dimers on spherical random lattices,''
  Phys.\ Lett.\ A {\bf 377}, 501 (2013),
  arXiv:1201.1931. 
    
\bibitem{Bonzom:2012qx} 
  V.~Bonzom and H.~Erbin,
  ``Coupling of hard dimers to dynamical lattices via random tensors,''
  J.\ Stat.\ Mech.\  {\bf 1209}, P09009 (2012),
  arXiv:1204.3798.

\bibitem{spinglass} 
M. M\'ezard, G. Parisi, and M. Virasoro, ``Spin Glass Theory and Beyond," World Scientific, 1986.

\bibitem{GrossMezard} 
D. Gross and M. M\'ezard, ``The simplest spin glass,Ó Nucl. Phys. B 240 no. 4, (1984) 431-452.

\bibitem{BGSspinglass} 
V. Bonzom, R. Gurau and  M. Smerlak,
``Universality in p-spin glasses with correlated disorder",
J. Stat. Mech. (2013) L02003,   arXiv:1206.5539.


\bibitem{DiFrancesco:1999tp} 
  P.~Di Francesco, O.~Golinelli and E.~Guitter,
  ``Meanders: Exact asymptotics,''
  Nucl.\ Phys.\ B {\bf 570}, 699 (2000),
cond-mat/9910453.

\bibitem{Grosse:2004yu}
  H.~Grosse and R.~Wulkenhaar,
``Renormalisation of phi**4 theory on noncommutative R**4 in the matrix base,''
  Commun.\ Math.\ Phys.\  {\bf 256}, 305 (2005),
   arXiv:hep-th/0401128.
  
\bibitem{Gurau:2008vd}
  R.~Gurau, J.~Magnen, V.~Rivasseau and A.~Tanasa,
  ``A translation-invariant renormalizable non-commutative scalar model,''
  Commun.\ Math.\ Phys.\  {\bf 287}, 275 (2009),
  arXiv:0802.0791.
  
\bibitem{Eichhorn:2013isa} 
  A.~Eichhorn and T.~Koslowski,
  ``Continuum limit in matrix models for quantum gravity from the Functional Renormalization Group,''
  arXiv:1309.1690. 
  
\bibitem{Grosse:2004by} 
  H.~Grosse and R.~Wulkenhaar,
  ``The beta function in duality covariant noncommutative phi**4 theory,''
  Eur.\ Phys.\ J.\ C {\bf 35}, 277 (2004),
  hep-th/0402093.
   
\bibitem{Disertori:2006uy} 
  M.~Disertori and V.~Rivasseau,
  ``Two and three loops beta function of non commutative Phi(4)**4 theory,''
  Eur.\ Phys.\ J.\ C {\bf 50}, 661 (2007),
  hep-th/0610224.
  
\bibitem{Disertori:2006nq} 
  M.~Disertori, R.~Gurau, J.~Magnen and V.~Rivasseau,
  ``Vanishing of Beta Function of Non Commutative Phi**4(4) Theory to all orders,''
  Phys.\ Lett.\ B {\bf 649}, 95 (2007),
  hep-th/0612251.
  
\bibitem{Grosse:2009pa} 
  H.~Grosse and R.~Wulkenhaar,
  ``Progress in solving a noncommutative quantum field theory in four dimensions,''
  arXiv:0909.1389. 
  
\bibitem{Grosse:2012uv} 
  H.~Grosse and R.~Wulkenhaar,
  ``Self-dual noncommutative $\phi^4$-theory in four dimensions is a non-perturbatively solvable and non-trivial quantum field theory,''
  arXiv:1205.0465. 
  
\bibitem{Geloun:2012bz} 
  J.~Ben Geloun and E.~R.~Livine,
  ``Some classes of renormalizable tensor models,''
  J.\ Math.\ Phys.\  {\bf 54}, 082303 (2013),
  arXiv:1207.0416. 
 
\bibitem{Krajewski:2012is} 
  T.~Krajewski, V.~Rivasseau and A.~Tanasa,
  ``Combinatorial Hopf algebraic description of the multiscale renormalization in quantum field theory,''
  arXiv:1211.4429.
  
\bibitem{Raasakka:2013kaa} 
  M.~Raasakka and A.~Tanasa,
  ``Combinatorial Hopf algebra for the Ben Geloun-Rivasseau tensor field theory,''
  arXiv:1306.1022. 

\bibitem{Carrozza:2013mna} 
  S.~Carrozza,
  ``Tensorial methods and renormalization in Group Field Theories,''
  arXiv:1310.3736. 

\bibitem{BenGeloun:2011xu} 
  J.~Ben Geloun,
``Ward-Takahashi identities for the colored Boulatov model,''
  J.\ Phys.\ A {\bf 44}, 415402 (2011),
  arXiv:1106.1847. 
    
\bibitem{Geloun:2013zka} 
  J.~B.~Geloun,
  ``On the finite amplitudes for open graphs in Abelian dynamical colored Boulatov-Ooguri models,''
  J.\ Phys.\ A {\bf 46}, 402002 (2013),
  arXiv:1307.8299. 


    
\bibitem{Oriti:2013jga} 
D.~Oriti,  ``Disappearance and emergence of space and time in quantum gravity,''
  arXiv:1302.2849 [physics.hist-ph].

\bibitem{Gielen:2013kla}
  S.~Gielen, D.~Oriti and L.~Sindoni,
  ``Cosmology from Group Field Theory Formalism for Quantum Gravity,''
  Phys.\ Rev.\ Lett.\  {\bf 111} (2013) 031301,
  arXiv:1303.3576. 


\bibitem{OS} K. Osterwalder and R. Schrader, ``Axioms for Euclidean GreenÕs functions", Comm.
Math. Phys. 31 (1973), 83Ð112 and 42 (1975) 281Ð305.

\bibitem{GRS} F. Guerra, L. Rosen and B. Simon, ``The $P(\phi)_2$
Euclidean quantum field theory as classical statistical
mechanics", Ann. Math. 101, 111 (1975).

\bibitem{DGR} 
T. Delepouve, R. Gurau and V. Rivasseau, in preparation.
 
\end{thebibliography}
\end{document}